\documentclass[twocolumn,twocolappendix]{aastex63}
\usepackage{amsmath}
\usepackage{CJK}

\newcommand\kms{km s$^{-1}$}

\newcommand\teff{$T_{\rm eff}$}
\newcommand\logg{$\log g$}

\usepackage{listings}

\begin{document}
\begin{CJK*}{UTF8}{gbsn}
\title{A self-consistent data-driven model for determining stellar parameters from optical and near-IR spectra}

\author{Logan Sizemore}
\affil{Computer Science Department, Western Washington University, 516 High St, Bellingham, WA 98225, USA}
\author{Diego Llanes}
\affil{Computer Science Department, Western Washington University, 516 High St, Bellingham, WA 98225, USA}
\author[0000-0002-5365-1267]{Marina Kounkel}
\affil{Department of Physics and Astronomy, University of North Florida, 1 UNF Dr., Jacksonville, FL, 32224}
\affil{Department of Physics and Astronomy, Vanderbilt University, VU Station 1807, Nashville, TN 37235, USA}
\author[0000-0002-5537-008X]{Brian Hutchinson}
\affil{Computer Science Department, Western Washington University, 516 High St, Bellingham, WA 98225, USA}
\affil{Foundational Data Science Group, Pacific Northwest National Laboratory, 902 Battelle Boulevard Richland, WA 99354, USA}
\author{Keivan G.\ Stassun}
\affil{Department of Physics and Astronomy, Vanderbilt University, VU Station 1807, Nashville, TN 37235, USA}
\author[0000-0002-0572-8012]{Vedant Chandra}
\affil{Center for Astrophysics $|$ Harvard \& Smithsonian, 60 Garden St., Cambridge, MA 02138, USA}
\email{marina.kounkel@unf.edu}

\begin{abstract}

Data-driven models, which apply machine learning to infer physical properties from large quantities of data,
have 
become increasingly important 
for extracting stellar properties from 
spectra. 
In general, these methods have been applied to data in one wavelength regime or another.
For example, APOGEE Net 
has been applied to near-IR spectra from the SDSS-V APOGEE survey to predict stellar parameters (Teff, log g, and [Fe/H]) for all stars with Teff from 3,000 to 50,000 K, including pre-main sequence stars, OB stars, main sequence dwarfs, and red giants. 
The 
increasing number of large surveys across multiple wavelength regimes provides the opportunity to improve data-driven models through learning from multiple datasets at once.
In SDSS-V, a number of spectra of stars will be observed not just with APOGEE in near-IR, but also with BOSS in optical regime. 
Here we aim to develop a complementary model, BOSS Net, that will replicate the performance of APOGEE Net in these optical data through label transfer. 
We further improve the model by extending it to brown dwarfs, as well as white dwarfs, resulting in a comprehensive coverage between 1700$<$\teff$<$100,000 K and 0$<$\logg$<$10, to ensure BOSS Net can reliably measure parameters of most of the commonly observed objects within this parameter space.
We also update APOGEE Net to achieve a comparable performance in the near-IR regime. The resulting models provide a robust tool for measuring stellar evolutionary states, and in turn, enable characterization of the star forming history of the Galaxy. \\
\end{abstract}


\section{Introduction}

Sloan Digital Sky survey in its fifth iteration (SDSS-V) is aiming to obtain spectra of several million stars across the Galaxy covering a wide range of ages and masses \citep{kollmeier2017}. This necessitates
 efficiently and homogeneously deriving their stellar properties to enable subsequent analyses.

Although a number of pipelines exist to derive parameters of stars, usually through comparing spectra to theoretical templates \citep[e.g.,][]{garcia-perez2016}, such approaches usually have significant limitations. These templates do not always accurately describe the data, and as a result they produce a number of systematic features that complicate interpretation of the data. This affects certain types of stars to a greater extent than others, as models can not always incorporate complex features of e.g., late type stars in full \citep{cottaar2014,kounkel2018a}. More boutique types of data processing pipelines also exist, but they are most efficient when focusing on a narrow parameter space \citep[e.g.,][]{souto2022}.

Data driven pipelines are an alternative approach. They make it possible to generalize across several boutique solutions for different stars of different types of stellar types in order to create a more self consistent solution that may improve on the original.

A number of data driven pipelines have been developed for SDSS data products, each with a somewhat different approach, incrementally improving on its predecessor through expanding the resulting parameter space. The Cannon \citep{ness2015} primarily focused its efforts on red giants. The Payne \citep{ting2019} has improved processing of \teff, \logg, and abundances of the solar type main sequence stars. APOGEE Net has incorporated pre-main sequence stars and K and early M dwarfs \citep{olney2020} as well as OB stars \citep{sprague2022} into the mix as well. However, the bulk of these data has been specifically developed for the spectra from APOGEE, which is the instrument that has conducted the bulk of the observations of stellar objects in the previous iterations of the survey.

In addition to APOGEE, SDSS also utilizes a second spectrograph, BOSS. Previously, BOSS has primarily been used to observe extragalactic objects, with only a sparse stellar program \citep[e.g.,][]{yan2019a,imig2022}. In SDSS-V, the scope of stellar observations with BOSS has been significantly expanded \citep{almeida2023}: in the first year of operations alone it has increased the number of stars for which it has obtained a spectrum by an order of magnitude compared to the entirety of SDSS-IV. In some cases, both the APOGEE and BOSS spectra can be obtained for the same stars, thus making it beneficial to produce a self-consistent solution for both instruments.

In this paper we present a data-driven pipeline, BOSS Net, that takes advantage of previous efforts to measure stellar parameters across different surveys in order to create a model for characterization of \teff, \logg\, and [Fe/H] in optical (BOSS \& LAMOST) spectra. We also provide an update to APOGEE Net to take advantage of newly available training sets. In Section \ref{sec:data} we describe the data used and the manner in which the labels have been derived for training the pipeline. In Section \ref{sec:model} we describe the neural network model. In Section \ref{sec:results} we present the resulting parameters, and discuss them in Section \ref{sec:discussion}. We conclude in Section \ref{sec:conclusions}.

\section{Data} \label{sec:data}

\subsection{Spectra}

Baryon Oscillation Spectroscopic Survey (BOSS) is an optical spectrograph covering a wavelength range of 3622--10354  \AA\ with the resolution of $R\sim1800$ \citep{smee2013}. Apache Point Observatory Galactic Evolution Experiment (APOGEE) is a near-IR specgrograph covering a range of 1.51--1.7 $\mu$m with $R\sim22,500$ \citep{wilson2010,majewski2017,wilson2019}. Both of them are installed at the Apache Point Observatory 2.5m telescope \citep[APO,][]{gunn2006,blanton2017}, and which is capable of obtaining up to 500 BOSS and 300 APOGEE spectra simultaneously in a given field \citep{pogge2020}. A similar set up has also been mounted at the Las Campanas Observatory DuPont 2.5m telescope \citep[LCO,][]{bowen1973}; combined they offer a full view of the entire sky. APO has the field of view of 3$^\circ$ with the fiber diameter of 2'', LCO has the field of view of 2$^\circ$, with the fiber diameter of 1.3''. BOSS can position its fibers through a robotic positioned, enabling rapid reconfiguration, and it operates simultaneously with the APOGEE spectrograph.

To date, BOSS has obtained more than 500,000 stellar spectra of $>$300,000 objects, rapidly increasing their census by the night. Despite this, so far there is not a significant overlap between BOSS \& APOGEE observations. In part, this is due to initial limitations of the bright limit of BOSS, being able to only observe stars fainter than $G>$13 mag. Strategies for overcoming this limitation are currently in place, which would allow to have more spectra in common between two instruments. In part, however, this can also be attributed to the targeting strategy: as BOSS can observe fainter stars, it includes many targets that are inaccessible to APOGEE in the first place, including white and brown dwarfs.

The limited overlap makes it difficult to transfer labels for \teff, \logg, and [Fe/H] from APOGEE to BOSS directly. However, it can be done through an intermediate step.

Large Sky Area Multi-Object Fibre Spectroscopic Telescope (LAMOST) is a spectrograph operated by the National Astronomical Observatories, Chinese Academy of Sciences. It has $R\sim1800$, covering the wavelength range of 3700--9000 \AA, capable of obtaining up to 4000 spectra in a single exposure \citep{yan2022}. Having been operated for over a decade, it has obtained spectra of 10 million stars to date, and it has 100,000s of stars in common with APOGEE, making it possible to transfer the labels. In addition, a number of pipelines have been developed for LAMOST specifically to characterize different corners of the parameter space \citep[e.g.,][]{lee2015,ho2017,du2021}, although, so far there has not been a comprehensive pipeline that is able to take advantage of all of its data in a self-consistent manner.

In many respects, with exception of the wavelength coverage, LAMOST is a very similar instrument to BOSS, down to the reduction pipelines utilized by the two surveys, and the resulting data architecture. As such, a data processing pipeline can be built to operate on spectra of both instruments. This enables different sets of labels both for BOSS \& LAMOST spectra to be used in a complementary manner in constructing the necessary model for their characterization.

It should be noted that despite similarity between the two instruments, there is some difference between the SNR of these two respective datasets. Since transitioning to the fiber positioner system, SDSS has limited the exposure time to only 15 min. BOSS has also been unable to observe stars brighter than $\sim G_{\rm RP}$<12 mag due to the saturation point of the instrument. As such, very few sources have SNR$>30$. On the other hand, SNR of $>100$ is not uncommon in LAMOST.

\subsection{Stellar Parameters}

To construct the training set of sources with previously measured \teff, \logg, and [Fe/H], we utilized a variety of approaches. Here we describe the datasets used for training BOSS Net. The comparison between these datasets when multiple labels for same stars are available is discussed in Appendix \ref{sec:overview} Updates to the APOGEE Net training set are presented in Appendix \ref{sec:apogee}

\subsubsection{FGK stars}

A number of red giants and solar type main sequence stars observed by APOGEE have previously had their parameters estimated using The Payne \citep{ting2019}, which is a neural net trained on synthetic spectra. This catalog includes 1676 stars observed by BOSS and 73,403 stars observed by LAMOST. It is complemented by the sample produced by APOGEE Net \citep{olney2020,sprague2022}, which in part has been built on The Payne, expanding it into a greater range of the parameter spaces. The Payne has processed spectra only through SDSS DR14 \citep{abolfathi2018}, APOGEE Net includes all the spectra released through SDSS DR17 \citep{abdurrouf2022}, as such the latter includes a greater number of recently observed sources. APOGEE Net provides parameters to additional 441 stars observed by BOSS, and 92,016 stars observed by LAMOST that, alongside with the Payne parameters, we adopt for the training set.

However, while sources previously observed and characterized by APOGEE offer a good starting place for the initial model, it does have several blind spots. Training a model without filling in this gap biases the resulting predictions. Recently released Gaia DR3 includes estimates of \teff, \logg, and [Fe/H] from its spectra \citep{fouesneau2022}. Cross-matching against this catalog shows that sources in SDSS-IV APOGEE have systematic deficit in sources with \logg$>$4.4 at \teff$\sim$6000 K in comparison to a number of sources observed in SDSS-V due to the previous targeting strategy of the survey.

Because of this, we adopt Gaia-derived parameters for stars with 5500$<$\teff$<$7500 K, excluding those found on the red giant branch with \teff$<$6300 K and \logg$<$3.8. We also exclude all of the sources targeted by the APOGEE \& BOSS Young Star Survey \citep[ABYSS,][]{kounkel2023}, as Gaia does not accurately derive their spectroscopic parameters. 

We also adopt Gaia parameters for sources with 3800$<$\teff$<$6000 K and 3.8$<$\logg$<$5, limiting the sample to the programs within SDSS-V that have specifically targeted evolved stars with a cleanly defined main sequence. Combined, parameters for 57,763 LAMOST and 5903 BOSS spectra were obtained from Gaia; the number of sources was limited so as to prevent this parameter space from overwhelming the rest of the sample.

To further improve the [Fe/H]$<$-1 sample, we have added sources in the halo that observed in SDSS-V, the spectra of which have been processed with MINESweeper \citep[Chandra, V. in prep]{cargile2020}.
We utilized the internal \texttt{mwmhalo\_clean\_rcat\_V0.07\_MSG} version of the catalog.
This produced labels of 19,160 BOSS spectra, and they were crossmatched with LAMOST to yield additional 1949 spectra.

\subsubsection{Pre-main sequence stars} \label{sec:yso}
One of the areas where APOGEE Net has significantly improved over The Payne is in its characterization of pre-main sequence stars. In particular, the resulting \logg s are sensitive to the age of a young star, enabling it to be used independently of photometry in characterizing star forming history of a given population \citep[e.g.,][]{kounkel2022b}. Nonetheless, the absolute calibration is imperfect: the shape of the resulting observed isochrones runs in parallel to the main sequence, which is inconsistent with the theoretical isochrones or the masses and radii measurements from the young eclipsing binaries. In part, this is due to the limited sample of young stars that have been available for training, as well as imprecise initial labels.

We improve on these efforts. In the initial iteration training BOSS Net we adopt APOGEE Net solutions for the young stars that were in common, and we then apply this model on BOSS and LAMOST data. The resulting \teff\ is well calibrated to the older stars, and it appears to match well the properties of the pre-main sequence stars \citep{kounkel2023a}. We thus adopt the resulting \teff\ for the low mass sources targeted by ABYSS with SDSS-V \citep[ABYSS,][]{kounkel2023}, but we renormalize other parameters. In particular, we estimate the ages of these stars using Sagitta \citep{mcbride2021}. We interpolate the combination of ages and \teff\ against MIST isorchones \citep{choi2016} in order to estimate \logg. The resulting \teff\ \& \logg\ have then been used in subsequent iterations of the BOSS Net model. This amounts to 2808 young stars in LAMOST and 1948 stars in BOSS.

We also improve on [Fe/H] determination for these stars. Previously, APOGEE Net has adopted solar metallicity for all pre-main sequence stars, as its training set has included only a handful of nearby star forming regions that all appear to be chemically homogeneous: star forming regions that are spread out over $\sim500$ pc, and even more evolved young clusters such as Pleiades all have [Fe/H] to be almost precisely solar \citep{soderblom2009,spina2017,kos2021}.. With inclusion of a larger number of populations across the Galaxy, this approximation is no longer appropriate. Additionally, as the special treatment of the pre-main sequence stars was initially limited to the low mass stars, this created a gradient in [Fe/H] as a function of \teff\ \citep{roman-zuniga2023}.

There is a significant [Fe/H] gradient in the Galaxy as a function of radius $R$, reflective of the inside-out star forming history \citep[e.g.,][]{pilkington2012}. Young stars are expected have [Fe/H] most closely correlated with this gradient, as they did not have sufficient time to migrate from their birth sites, and thus act as excellent tracers of the chemical composition of the gas in which they formed. Thus, we adopt the relationship of 

\begin{equation}{\rm [Fe/H]}=-0.053R+0.428\end{equation}
from \citet{hayden2015}. At $R$=8 kpc, this does reproduce solar [Fe/H] of the nearby young populations, and the gradient is sufficiently shallow that at larger distances the uncertainties in parallax should have negligible effect. We adopt the resulting [Fe/H] for all ABYSS targets, regardless of their \teff; this was done for 2068 stars in BOSS and 4045 stars in LAMOST.

\subsubsection{OBA stars}

APOGEE Net has previously been enabled to estimate spectroscopic parameters of OBA stars. While the resulting \teff\ were reliable, and \logg\ did show some ability to differentiate between dwarfs and giants, it struggled to fully populate the entire \logg\ space that is expected by the theoretical models. This is in part due to the previous targeting strategy of the survey, lower sensitivity of \logg\ in NIR spectra, as well as imprecise labels -- typically obtained through interpolating the spectral type and luminosity class to \teff\ \& \logg. Furthermore, no [Fe/H] information was available. Although in the absence of other labels, we adopt APOGEE Net parameters, as well as parameters derived through a similar interpolation from the spectral type, it is necessary to compensate for this bias.

Since then, however, there has been a significant increase in the number of hot mass stars with accurately derived properties. In particular HotPayne \citep{xiang2022} has provided \teff, \logg, and abundances for a number of stars in LAMOST spectra. We adopt these parameters for sources in LAMOST spectra that reported precision in \teff$<$4\%, and precision in \logg\ of $<$0.2 dex. 

Since we begun this work, there has also been a separate release of the parameters of hot stars in SDSS-V data, zeta Payne \citep{straumit2022}. We did not include these labels, as the parameter distribution showed significantly larger number of systematic features compared to HotPayne.

The sample in HotPayne has a very sharp edge at 7000 K -- it may include sources that are somewhat cooler that got aliased towards $\sim7000$ K due to the edge effects. We cross-matched these sources with Gaia DR3 parameters, and have overwritten the parameters for the subset where Gaia has reported \teff\ between 6000 and 7000 K. 

Since the distribution of \teff\ in the full training set doesn't have a smooth distribution due to the inhomogeneity of the catalogs, we randomly downsampled the remaining hot stars very close to the boundary so that the transition between the sources in HotPayne and cool stars is more continuous. In total, our training of these sources consists of 128,430 LAMOST and 663 BOSS spectra. 

\subsubsection{Low mass stars \& brown dwarfs} \label{sec:mdwarfs}

\begin{figure}
\epsscale{1.1}
\plotone{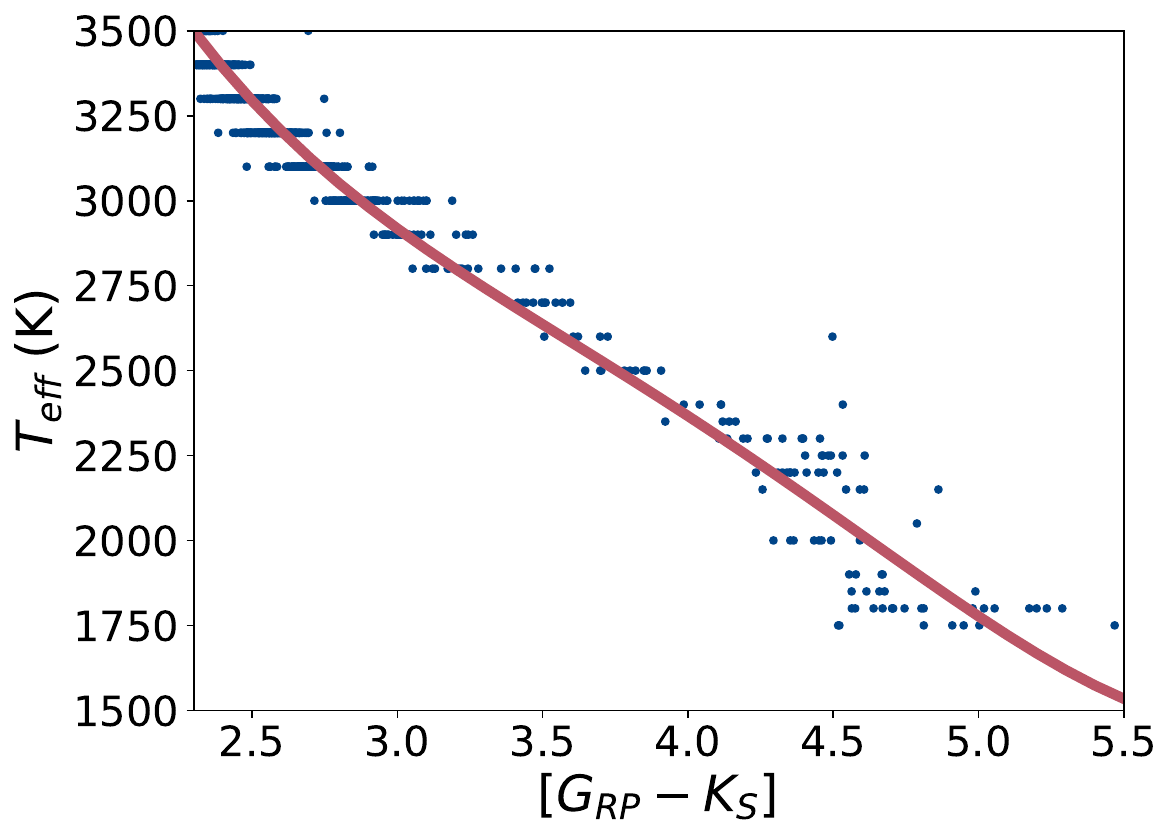}
\plotone{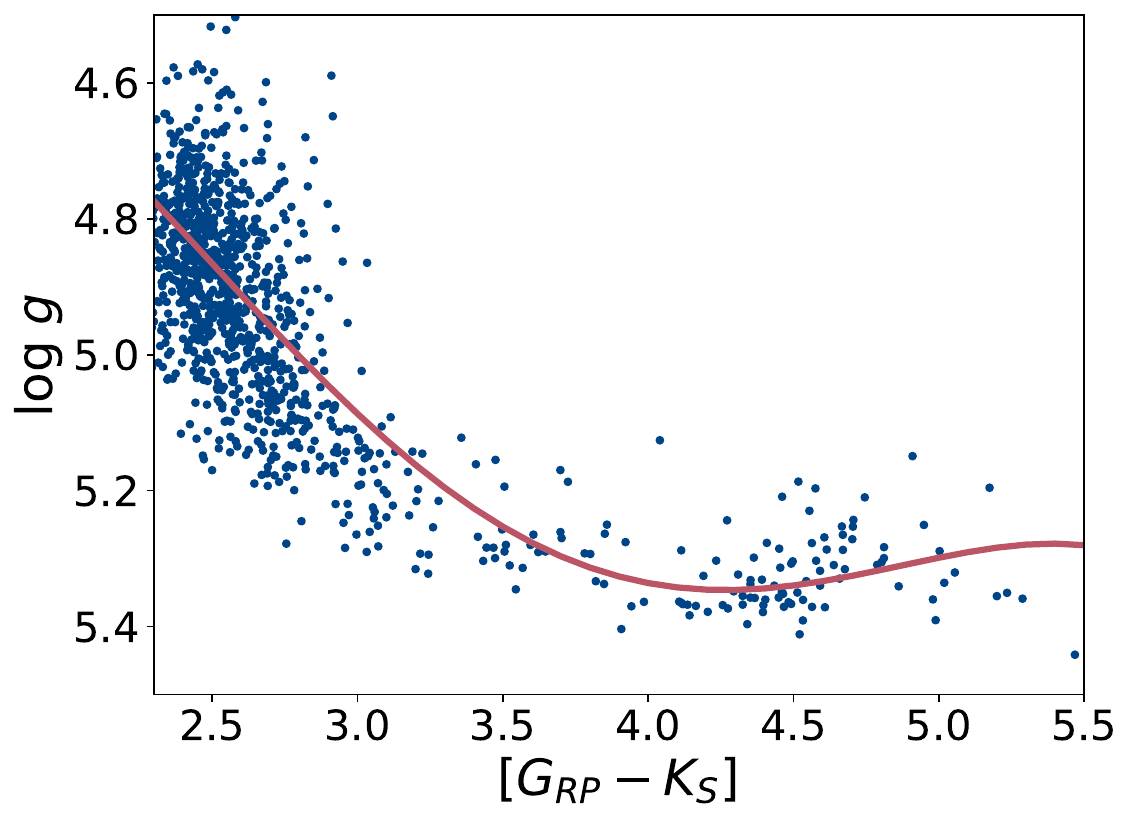}
\caption{Photometric interpolation of \teff\ and \logg\ values for cool dwarfs using the data from \citet{cifuentes2020}.
\label{fig:interp}}
\end{figure}

APOGEE Net is unable to provide stellar parameters to stars cooler than 3000 K, as, at the time it was initially developed, APOGEE did not generally obtain spectra of such objects. In the recent years, APOGEE has begun to observe these stars more routinely, and they are also common in BOSS spectra due to its higher sensitivity.

Since the goal of this project is to create a pipeline that is capable on running on all stellar optical spectra to provide homogeneous set of parameters without any unintended edge effects, we need to account for these cool stars. Unfortunately, at the moment of assembling the labels, have been very few reliable stellar parameters for these stars. Instead, to incorporate them into the model, we rely on photometric relations to derive their parameters (Figure \ref{fig:interp}). Specifically, we examine the properties of cool dwarfs from \citet{cifuentes2020} that were measured from the CARMENES spectra. We obtain an interpolation of:

\begin{equation}
\begin{split}
    T_{\rm eff}=12870-9027(G_{\rm RP}-K_S)+3201(G_{\rm RP}-K_S)^2\\-529.0(G_{\rm RP}-K_S)^3+32.25(G_{\rm RP}-K_S)^4
\end{split}
\end{equation}
\begin{equation}
\begin{split}
    \log g = 6.122-2.754(G_{\rm RP}-K_S)+1.546(G_{\rm RP}-K_S)^2\\-0.3109(G_{\rm RP}-K_S)^3+0.02105(G_{\rm RP}-K_S)^4
\end{split}
\end{equation}
that is valid for stars with $(G_{\rm RP}-K_S)>$2.3 mag, the scatter in the fit relative to the available data is 85 K in \teff, and 0.11 K in \logg. Unfortunately, no [Fe/H] information can be inferred from this sample.

We adopt parameters from this interpolation for 2853 BOSS and 9934 LAMOST spectra that satisfy $(G_{\rm RP}-K_S)>$2.3 mag, and $H-5\log 1000/\pi +5$>6 mag, and $\pi>10$ mas (for LAMOST) or targeted as a part of Solar Neighborhood Census (for BOSS) to select nearby faint and red stars that are most likely to be cool dwarfs.
.

\subsubsection{Subdwarfs \& white dwarfs}

Spectra of a number of compact objects have been observed both by LAMOST and APOGEE. Although they are substantively different from stars and are usually treated separately, their spectral features follow the same \teff\ and \logg relations as normal stars do. As such, including these objects in the training set enables developing a more self-consistent stellar model that can help compensating for imperfections in the labels of different classes.

We incorporate \teff\ and \logg\ parameters of 874 LAMOST and 12 BOSS spectra of hot subdwarfs from \citet{luo2019a}. A number of hot subdwarfs has also been incorporated in the catalog from HotPayne \citep{xiang2022}. We also include 2654 LAMOST and 218 SDSS-V BOSS spectra of white dwarfs (WD) from \citet{kepler2019}. More recently, \citet{gentile-fusillo2021} produced a more comprehensive catalog of spectral parameters of WDs in the legacy SDSS data, including not only hydrogen-rich DA types, but also helium-rich DB types, and several others. We adopt these parameters for 6821 legacy SDSS I-IV BOSS spectra, and 536 cross-matched LAMOST spectra. No [Fe/H] is available for these sources, although, given their evolved status, it is difficult to define this value on the same meaningful scale as in regular stars.

The catalogs of WDs from \citet{kepler2019} and \citet{gentile-fusillo2021} has only a few sources with \teff$<$7000 K. SDSS-V white dwarf selection function appears to include many cooler sources in its targeting. If they are not included in the training set, the model predictions produce a sharp transition of WD sequence onto the main sequence at \teff$\sim$7000 K, as the model doesn't recognize very cool sources with high surface gravity as a valid parameter space. In the preliminary training of the model, we identified of these cool sources: while their \logg\ was not accurate, their \teff\ was informed by the training set of stars. Following that preliminary training, we selected sources targeted as WDs that have landed onto the main sequence, and adapted them into the training set, preserving their \teff, but setting their \logg=8 (as their sizes should not change significantly as they cool down). Although the absolute calibration of \teff\ for these sources may systematically differ by a few 100s of K at these \logg, we expect that the overall ordering of sources for hottest to coldest to be sufficiently self-consistent. This was done for 1945 sources observed BOSS; LAMOST targeting appears to lack such sources.

\subsubsection{Summary}

\begin{figure}
\includegraphics[width={0.48\textwidth}]{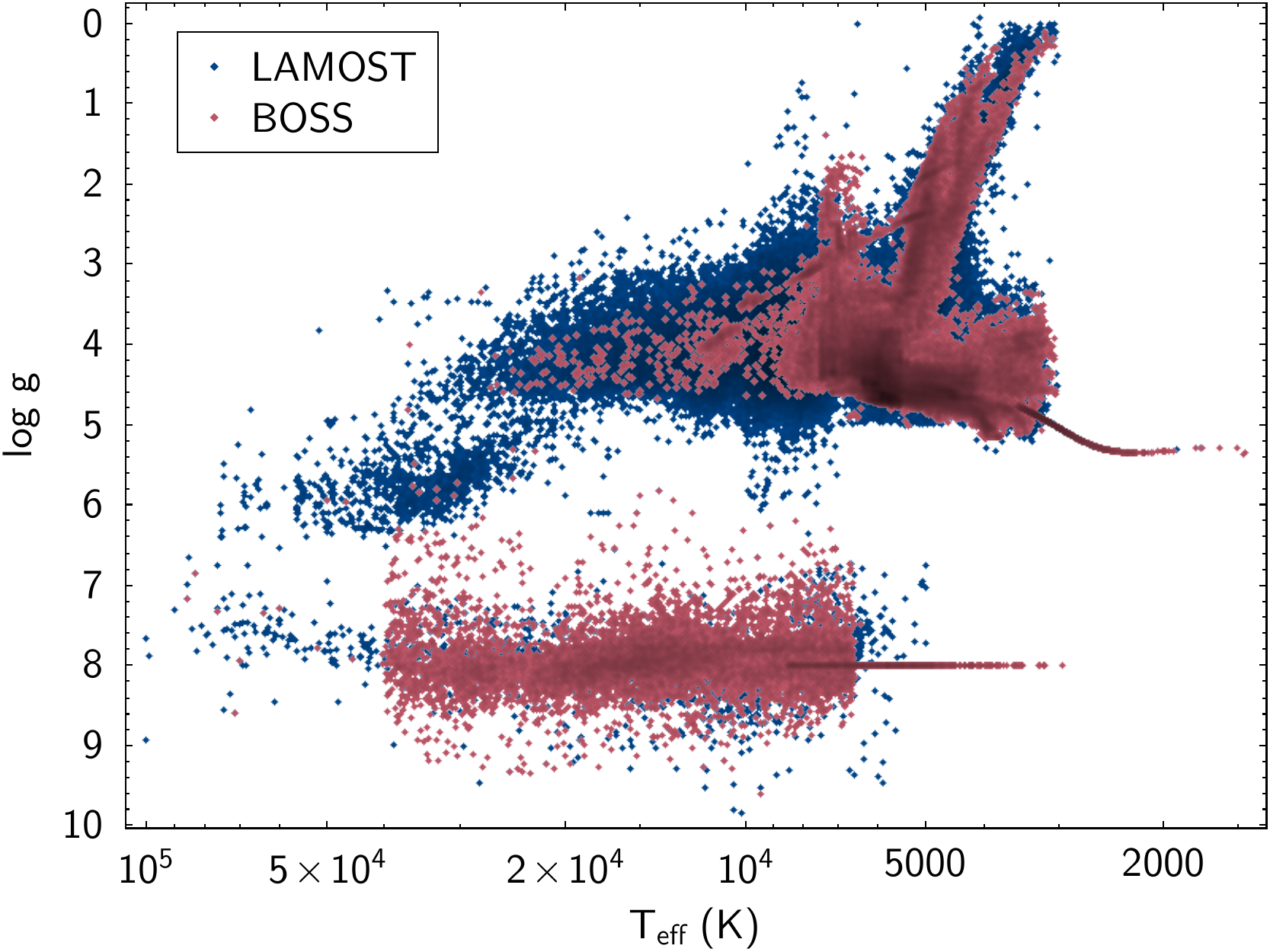}
\caption{Distribution of the \teff\ and \logg\ of the sources that form the training set, for BOSS in red, and LAMOST in blue.
\label{fig:training}}
\end{figure}

In total, the training set consists of 412,099 spectra, of which 371,072 in LAMOST, and 41,027 in BOSS. It covers 1,700$<$\teff$<$100,000 K and 0$<$\logg$<$10 (Figure \ref{fig:training}). Of them, 349,135 and 30,256 spectra respectively have label for [Fe/H], with 99\% of the sources occupying the range of -2$<$[Fe/H]$<$0.5; the sources with valid [Fe/H] occupy \logg$<$5 and \teff$>$3200 K.

\subsection{Radial velocity}

In the commissioning of SDSS-V, the wavelength solution of BOSS spectra has been rigorously tested to ensure self-consistency. Those tests have shown that the RVs derived using the pipeline from the previous generation of the survey can carry a significant offset of 5--10 \kms\ relative to high resolution spectra such as APOGEE. The performance in the red part of the detector was somewhat stable, but in the blue part, especially at $<$4000 \AA, it significantly degraded. Furthermore, this offset had a dependence on the pointing angle of the telescope, as it was not consistent across the sky. This amounts only to a fraction of a pixel on the detector, and the typical resolution of the instrument is only $\sim$5 \kms. Nonetheless, such an offset was considered a significant detriment to SDSS-V. Previously, BOSS had focused on extra-Galactic redshifts, for which such an offset is negligible.

With improved arc lamp line list, and better observing strategy for the calibration, the wavelength solution has subsequently stabilized. The pipeline PyXCSAO \citep{pyxcsao} was developed to improve the quality of the RVs, limiting the range of wavelengths over which cross-correlation was performed. The resulting RVs are consistent with RVs derived from APOGEE, and they are stable across all stars in various clusters (where all stars should have the same velocity), regardless of \teff\ \citep{kounkel2023}.

However, because of the limited wavelength coverage, while PyXCSAO is effective on cool stars, it does not function well for sources where the bulk of their light is in the blue portion of the spectrum. Among the sources for which it is unable to derive RVs are white dwarfs, which necessitates an alternate approach.

The issue of poor RVs is not unique to BOSS. Both LAMOST and BOSS pipelines share the common origins, and although both underwent significant modifications through the present day, issues with the wavelength solution appear to date back to SEGUE data \citep{yanny2009}, which was the original survey on which both of them were based. It is unclear to what degree these data can be reprocessed to improve RVs. As such, while PyXCSAO can offer some improvement in the RV stability over the native LAMOST pipeline, some systematic offsets remain, not in the least because the wavelength coverage of the instrument does not extend as far into the red.

As part of the stellar parameter determination, we aim to somewhat improve RV determination for both instruments. In our training set, we include stable RVs with $R>$6 reported by PyXCSAO for BOSS. We also cross-match the sources in the training set with APOGEE and other high resolution surveys collated in \citep{tsantaki2022}, to adopt RVs for some sources. While there are spectroscopic binaries which introduce RV variability, their overall fraction is expected to be $<$10\% of the total sample \citep{price-whelan2020}, of which significantly fewer would exhibit RV variability $>$5 \kms (the resolution of BOSS \& LAMOST), as such their presence should not significantly skew the model. We also adopt RVs from the most confident and stable determinations from Gaia DR3 \citep{katz2022}. For WDs from legacy SDSS spectra, we adopt RVs from \citep{anguiano2017}.

In total, RV data were available for 194,978 LAMOST spectra and 36,351 BOSS spectra that were already included in our training set.

\section{Model Description} \label{sec:model}

\subsection{Data processing}

\begin{figure*}
\epsscale{1.1}
\plotone{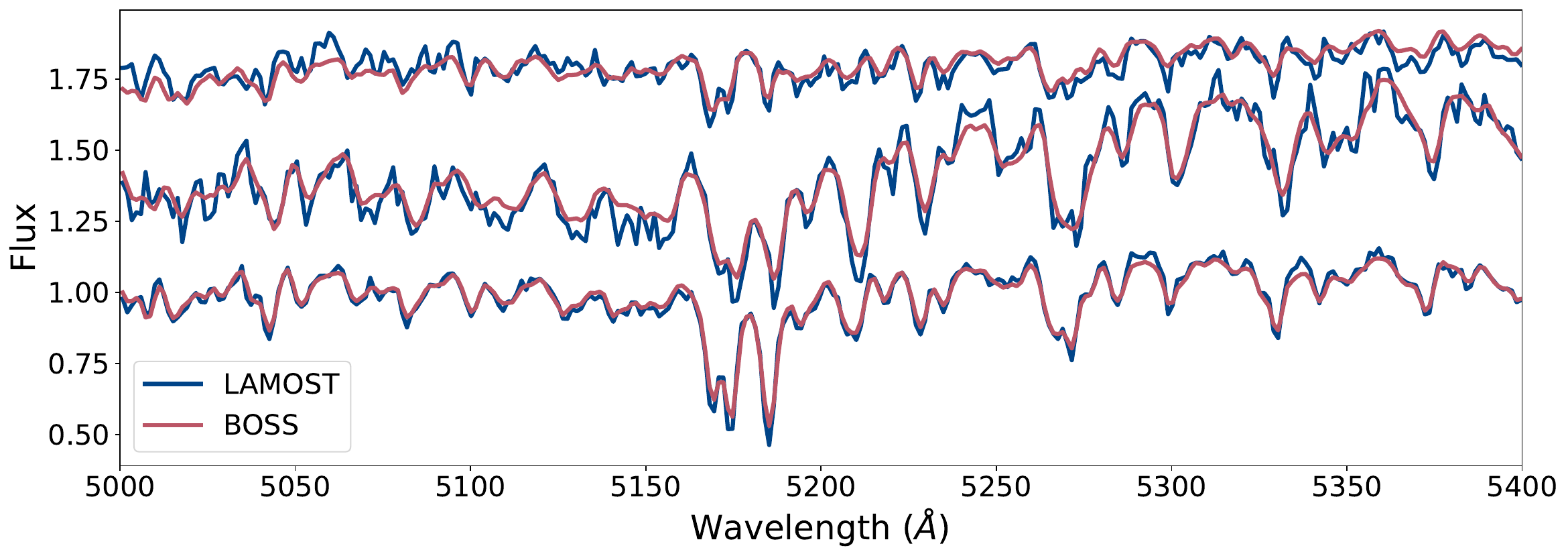}
\caption{Zoomed in comparison of BOSS and LAMOST spectra for some of the sources for which spectra from both instruments is available. All spectra have been interpolated onto a common wavelength grid. The shown sources have SNR$\sim$100 for BOSS, and $\sim$30 for LAMOST. Both BOSS and LAMOST spectra were normalized through dividing out the median flux within the presented wavelength range, and different sources were arbitrarily offset.
\label{fig:bosslamost}}
\end{figure*}

BOSS and LAMOST data have a comparable resolution (Figure \ref{fig:bosslamost}), and they cover a similar wavelength range, but they are not identical. Furthermore, since both of them are multi-object spectrographs, a given spectrum can be somewhat offset in wavelength depending on where it is positioned on a detector. In training of a model, this is not optimal, as the data need to be standardized with a consistent shape. To do this, we interpolate all of the data onto a common 3900 element wavelength grid ranging uniformly from 3800 to 8900 \AA, approximating the typical resolution of the data within that range. Because of this standardization, BOSS Net may have applications outside of just these two datasets and may function as a general tool for the optical spectra in general.

To compress the dynamical range of the data in order to improve the performance of the model and ensure numerical stability, we apply the log scaling to the flux, however, we do not perform continuum normalization, as it is non-trivial to do it self-consistently across all types of stars over such a large wavelength regime. The dataset comprising of LAMOST and BOSS stars was randomly segregated into distinct train, validation, and test sets, at a 80:10:10 ratio, respectively.% for training, validation, and testing respectively.

%\begin{deluxetable}{cccc}
%\tablecaption{Normalization constants used in training
%\label{tab:norm}}
%\tabletypesize{\scriptsize}
%\tablewidth{\linewidth}
%\tablehead{
%  \colhead{Parameter} &
%  \colhead{Mean} &
%  \colhead{$\sigma$} &
%  \colhead{Default}
%  }
%\startdata
%$\log$ \teff& 3.8 & 0.1 & \\
%\logg & 3.9 & 0.8 & \\
%Fe/H & -0.4 & 0.5 & \\
%\enddata
%\end{deluxetable}

\subsection{Model architecture \& training}

BOSS Net is a 1D residual convolutional network consisting of a series of 1D convolutional Residual Network blocks and a final linear network for prediction.

The BOSS Net model takes a star's spectrum as input. To regularize and avoid overfitting, data augmentation techniques are employed, such as randomly removing continuous segments of flux, dropping specific values in the flux, and adding noise to the flux by scaling a normal distribution sample by the error.

The model starts with a single convolutional block that includes a 1D convolutional layer, batch normalization, an Exponential Linear Unit (ELU) activation function, and a 1D max pooling layer. Batch normalization helps to improve the speed and stability of the training process by normalizing the inputs to each layer, reducing covariate shift, which is the change in input distribution of model layers as the model trains \citep{batch_norm}. The ELU activation function enhances the performance and convergence speed of the model compared to other activation functions such as Rectified Linear Units (ReLU) \citep{ELU}. The max pooling layer helps reduce the number of parameters in the model, which can prevent overfitting and improve generalization to unseen data. Additionally, the model includes a positional encoding as a channel to the first convolutional layer of the first block, allowing for more effective capturing of local patterns in the spectra at a given wavelength.

The output from the initial convolutional block flows into the first of many Residual Neural Network (ResNet) blocks. Each block consists of two 1D convolutional layers, each with batch normalization and an ELU activation function. The residual connection in this block allows the network to learn the residual mapping from the input to the output rather than the complete mapping, which can help with the vanishing gradient problem during training. ResNets were designed to allow for easier flow of information and gradients throughout the network, enabling the training of deeper models \citep{ResNets}.

%The metadata inputs are first embedded with a small linear network, which consists of two fully connected layers with the Gaussian Error Linear Unit (GELU) activation function. The use of the GELU as an activation function in the metadata network enables non-linear transformations of the metadata information and serves as a stochastic regularizer, improving the model's ability to capture the relationship between the metadata and the target variables \citep{GELU}. Then, in each ResNet block, the metadata embedding is first processed by an a single linear layer to generate scale and shift parameters for the residual connection, which are then applied to the residual to modify it. Specifically, these parameters operate by scaling and shifting each channel of the residual channel. By incorporating this learned modification step, the model is better able to capture underlying patterns within the data, leading to improved predictive accuracy.

%The metadata conditioned residual connection is added to the output of the second convolutional layer before being passed through the final ELU activation function. If the number of output channels is different from the number of input channels, the residual connection is padded with zeros on either side of the channel axis to match the size of the output.

After the output of the residual blocks is obtained, it is passed through an adaptive average pooling layer. This layer serves to adjust the dimensions of the outputs to a fixed size, allowing for easier integration with the final linear network. The outputs of the adaptive pooling layer are then fed into several linear layers, each followed by the ReLU activation function. The final linear network provides the model prediction for the star's \teff, \logg, [Fe/H], and RV.

\begin{figure}
\epsscale{1.2}
\plotone{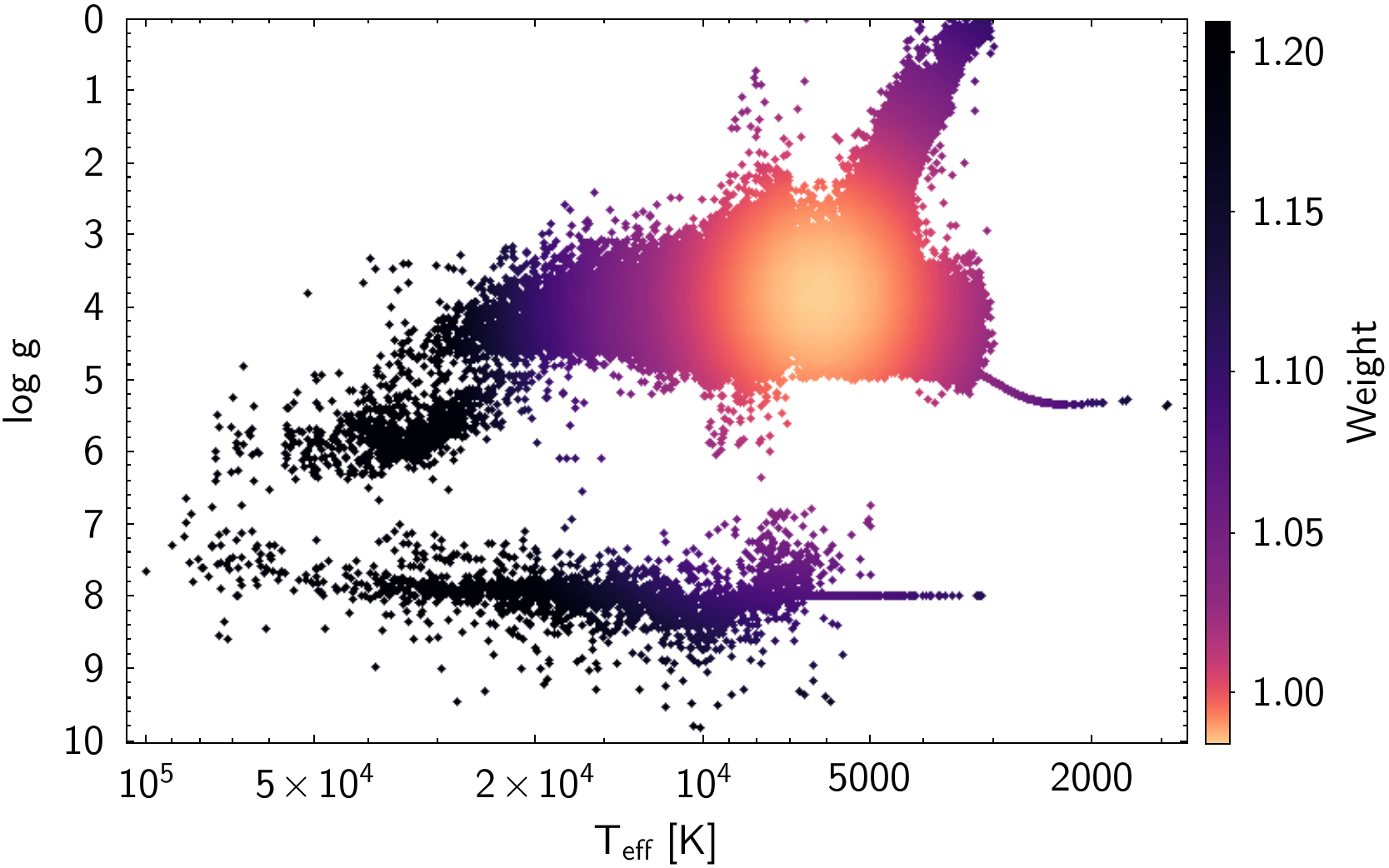}
\caption{Distribution of the \teff\ and \logg\ from the training set colored according to the weight applied to the loss.
\label{fig:kde_weights}}
\end{figure}

BOSS Net is evaluated with the mean squared error between the predicted and actual values. To encourage accurate modeling of the less populated regions of the parameter space, the loss of stars was adjusted, either by increasing or decreasing their weight in the loss calculation during training. The weight of specific regions is determined by the reciprocal of the Kernel Density Estimation (KDE) values computed from the training labels. This allows the model to prioritize the learning of less common samples, as indicated by their low-density regions in the KDE estimation, while reducing the impact of well-modeled samples, which are associated with high-density regions. As shown in \ref{fig:kde_weights}, the white dwarfs and the hot subdwarfs are weighted higher in the loss calculation, as they are in less dense regions of the parameter space. In addition, [Fe/H] of metal poor stars is weighted higher in comparison to the metal rich stars, and RVs of WDs or stars with |RV|$>$200 \kms\ are weighed higher than in other sources, once again due to their rarity in the sample. The model was trained with a learning rate of 0.0001 using the Adamax optimizer \citep[which is more robust to the presence of outliers,][]{kingma2014adam}, and a batch size of 512.

\begin{figure}
\includegraphics[trim={5cm 8cm 5cm 1.3cm},clip, width={0.45\textwidth}]{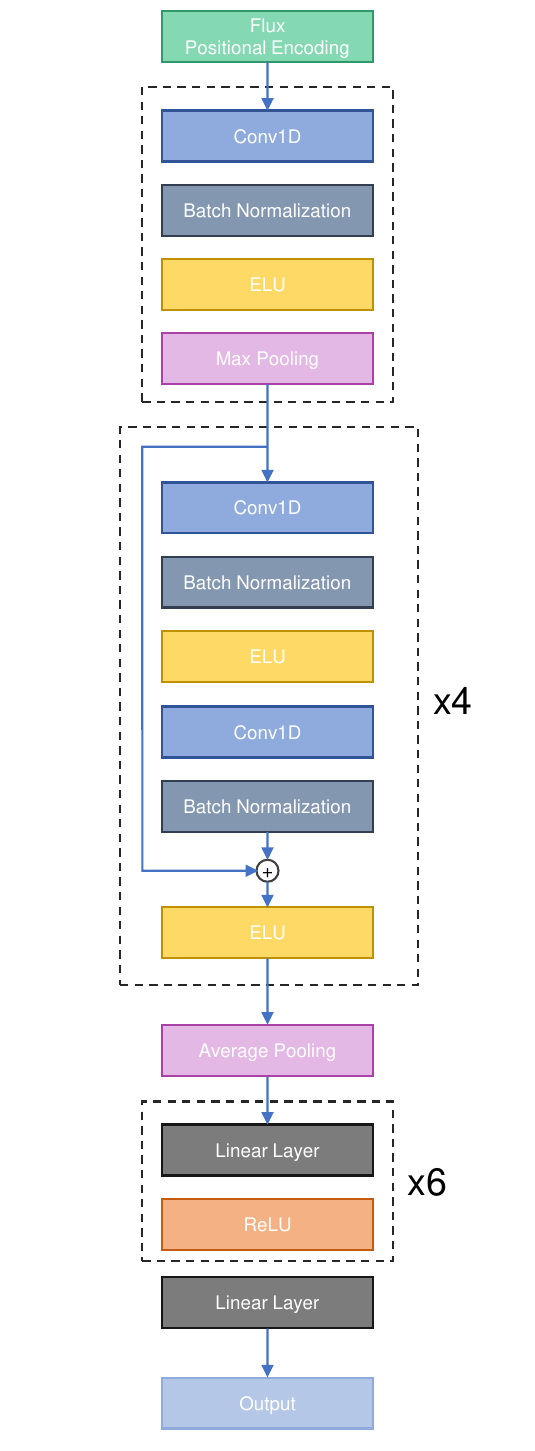}
\caption{Architecture of the neural net model used in this paper.
\label{fig:model}}
\end{figure}

\section{Results} \label{sec:results}

\begin{deluxetable}{ccl}[!ht]
\tablecaption{Stellar properties from LAMOST, Legacy SDSS spectra, and MaStar stellar library 
\label{tab:lamost}}
\tabletypesize{\scriptsize}
\tablewidth{\linewidth}
\tablehead{
 \colhead{Column} &
 \colhead{Unit} &
 \colhead{Description}
 }
\startdata
obsid & & LAMOST unique identifier \\
RA & deg & Right ascention in J2000 \\
Dec & deg & Declination in J2000 \\
log \teff & [K] & Effective temperature \\
$\sigma$ log \teff & [K] & uncertainty in log \teff \\
log g &  & Surface gravity \\
$\sigma$ log g & & Uncertainty in log g \\
$\left[\rm{Fe/H}\right]$&  & Metallicity \\
$\sigma$ [Fe/H] & & Uncertainty in [Fe/H] \\
RV & \kms & Radial velocity \\
$\sigma$ RV & \kms & Uncertainty in RV \\
\enddata
\end{deluxetable}

\begin{figure*}
\epsscale{1.1}
\plottwo{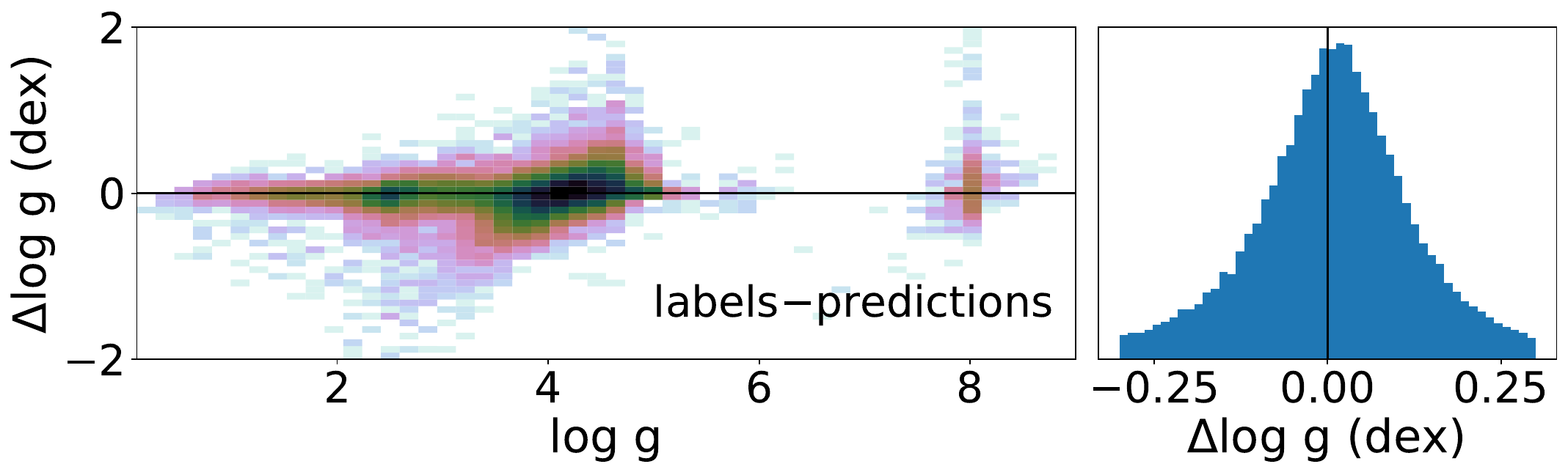}{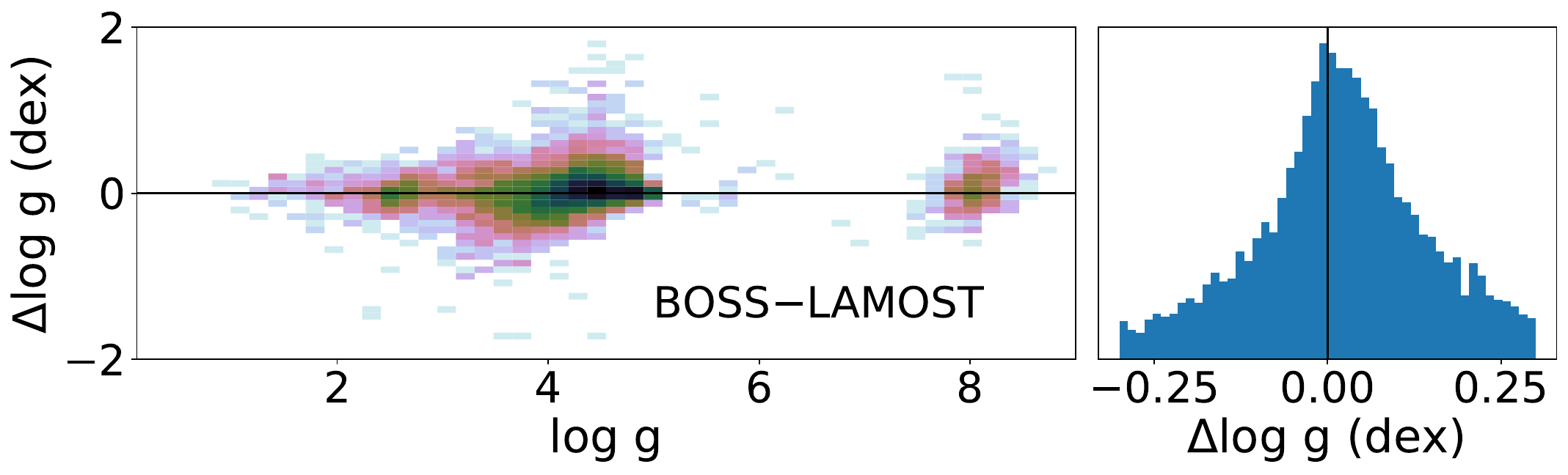}
\plottwo{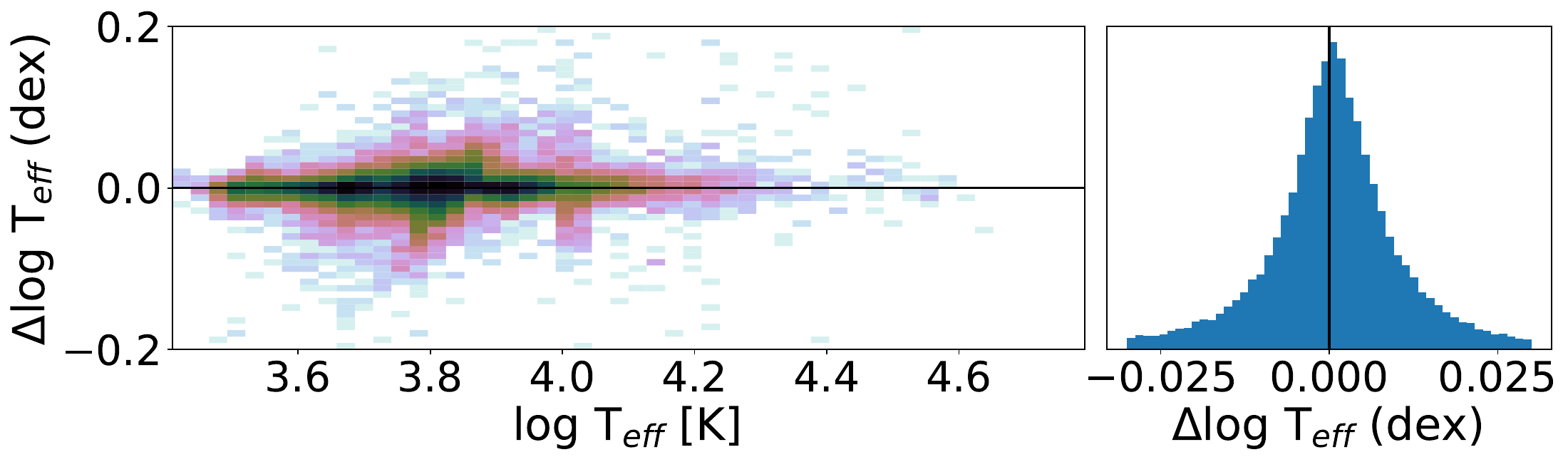}{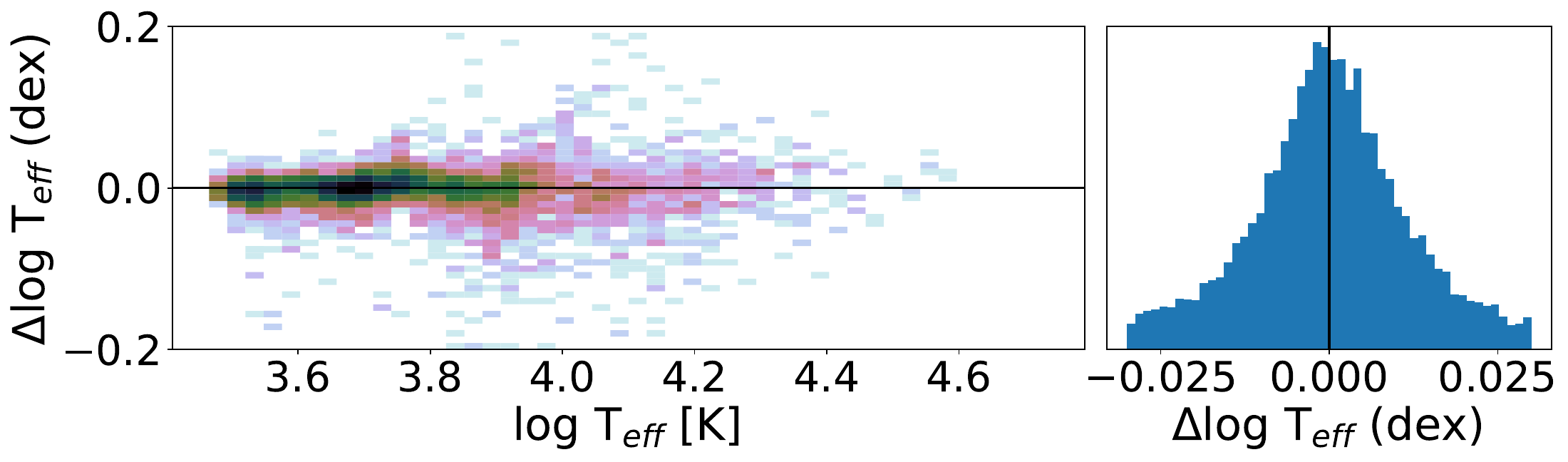}
\plottwo{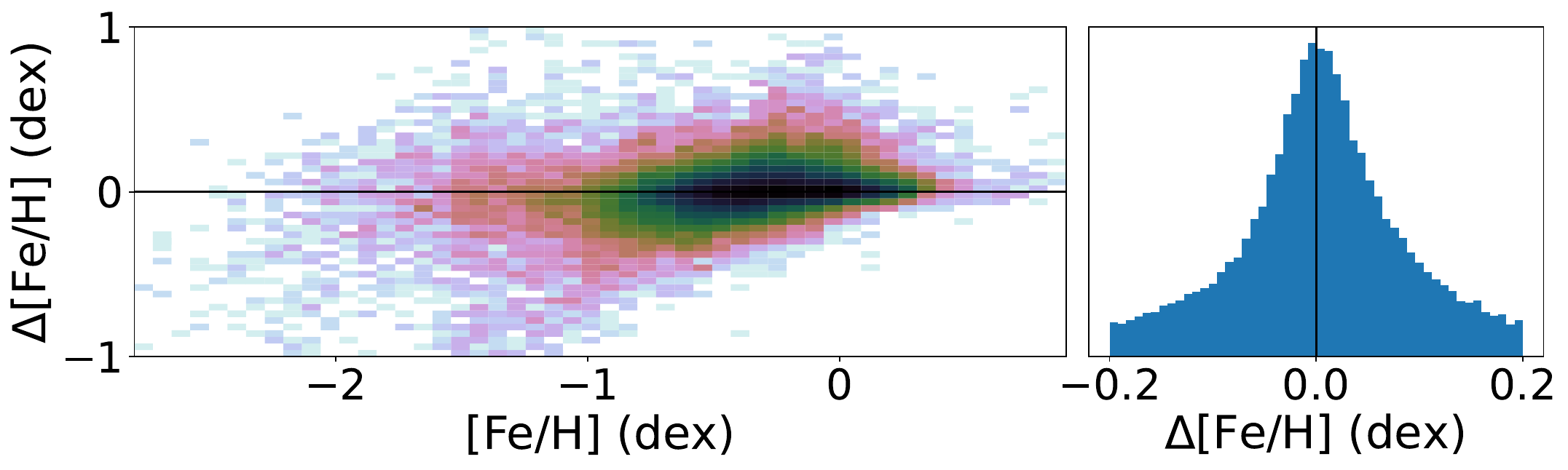}{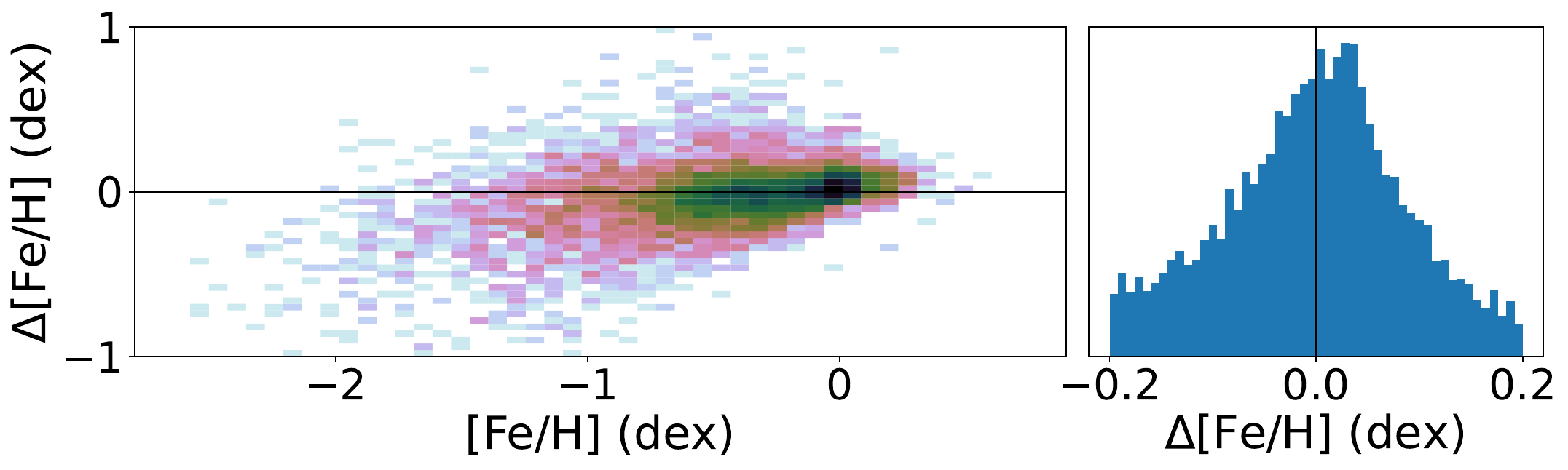}
\plottwo{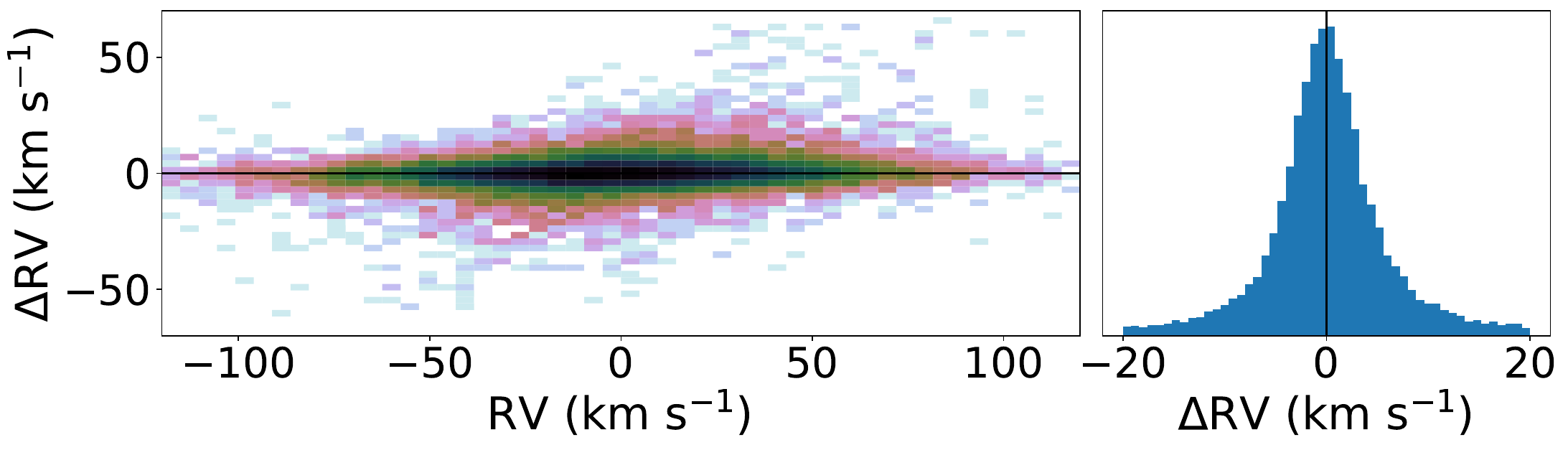}{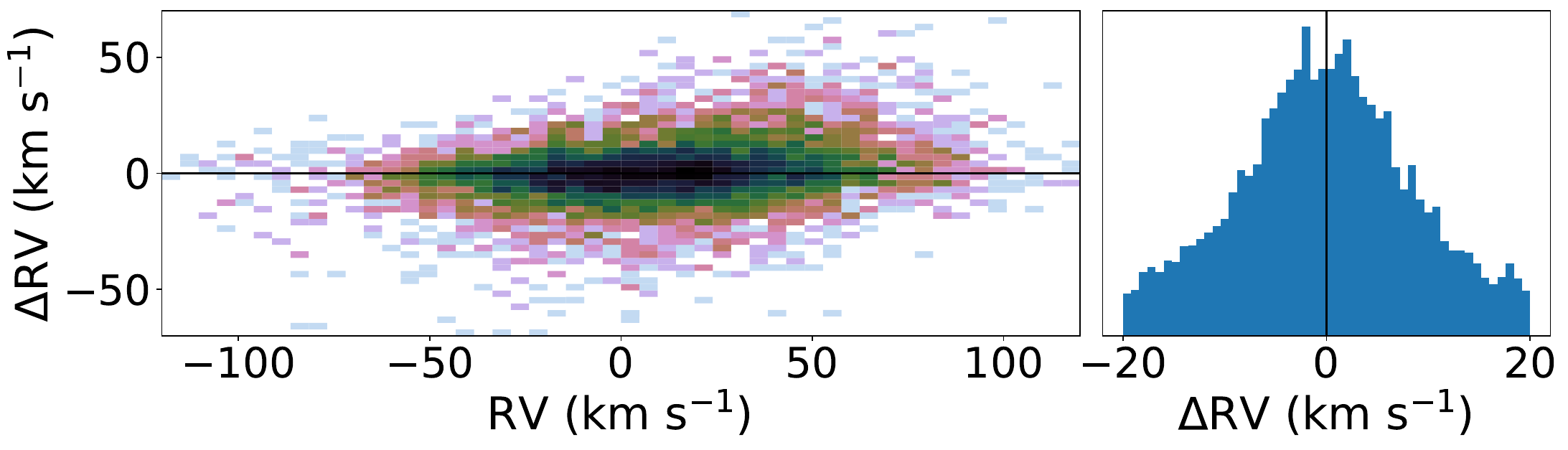}
\caption{Left: The performance of BOSS Net on the withheld test set, showing the consistency between the original labels for \teff, \logg, and [Fe/H] versus the resulting predictions. X-axis shows the labels. Right: Comparison between the derived parameters from BOSS \& LAMOST spectra of the sames sources. X-axis shows the BOSS measurements.
\label{fig:test1}}
\end{figure*}

\begin{figure*}
\epsscale{1.1}
\plotone{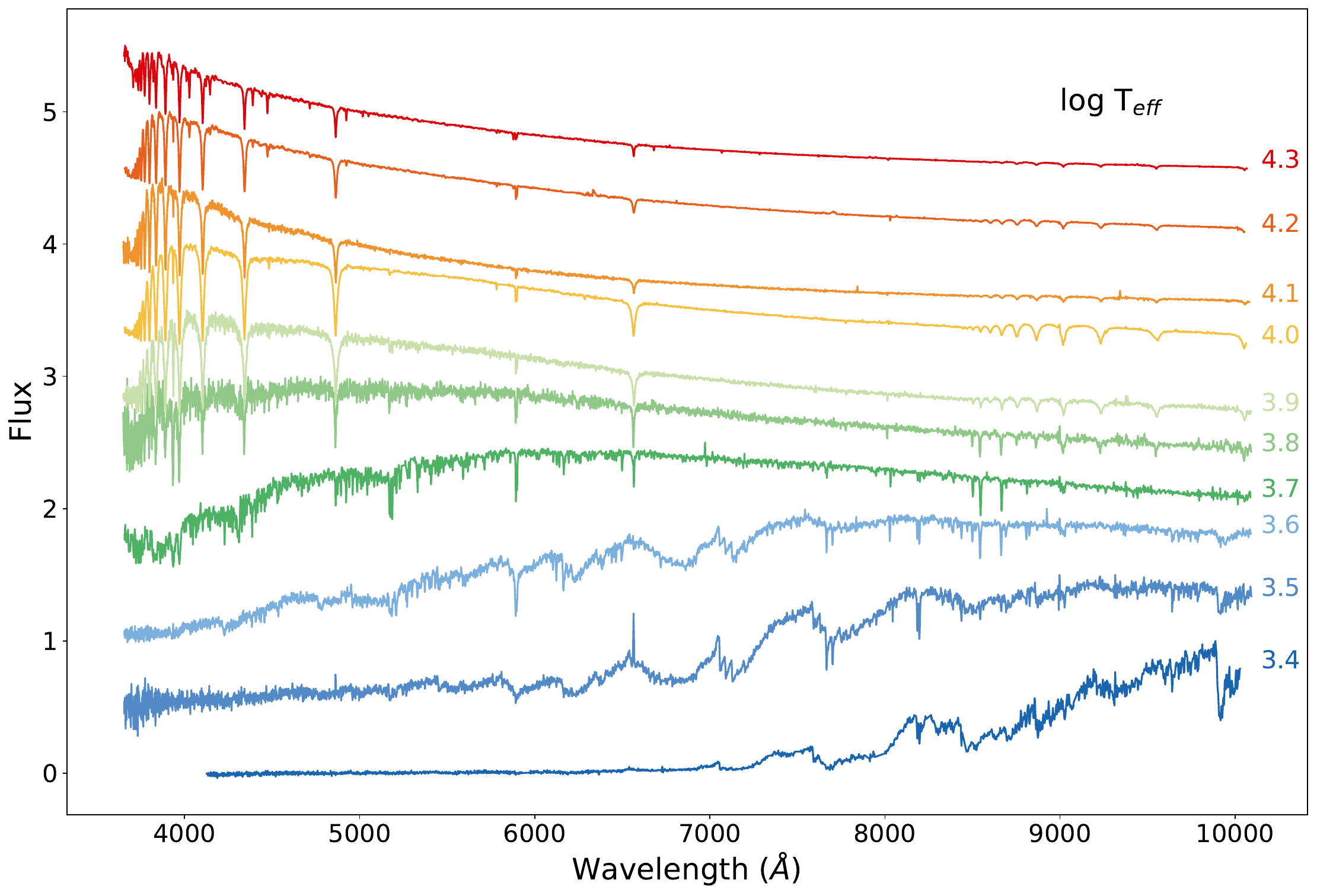}
\caption{Example BOSS spectra for stars of different \teff.
\label{fig:boss_teff_comp}}
\end{figure*}

In Table \ref{tab:lamost} we present BOSS Net derived parameters for LAMOST DR8 and stellar objects observed with BOSS through SDSS DR18, which also includes the legacy SEGUE data, as well as those in the MaStar library \citep{imig2022}. SDSS-V BOSS data are still proprietary to the collaboration. The code is integrated into Astra, the primary stellar data processing pipeline for SDSS-V. As such, its data products will be included in the subsequent data releases.

As the model was trained on the data in the training set, it was regularly evaluated against the validation set to ensure an adequate ability to generalize on the unseen data, and to prevent the model from overfitting. The training is terminated at the point when the performance on the validation set stops improving.

Since the validation set does end up influencing the model, once the model is finalized, it is evaluated against the test set. The resulting comparison between the predictions and the labels for these data are shown in Figure \ref{fig:test1}. The detail comparison of the predictions with respect to each subset of labeled data is available in Appendix \ref{sec:overview}.

In comparison to the initial labels, some of the cool white dwarfs may have their \logg\ underestimated, placed in between the main and white dwarf sequences; these tend to be lower SNR sources where confident spectroscopic determination of their \logg\ is difficult. Similarly, there are some sources with very metal poor labels, [Fe/H] of which is overestimated in the preditions; these tend to be hotter stars which intrinsically have very weak metal lines, and these labels typically originate from Gaia (near the edge of their [Fe/H] grid) which may not have as much sensitivity to [Fe/H] as higher resolution spectra.

\begin{figure}
\epsscale{1.2}
\plotone{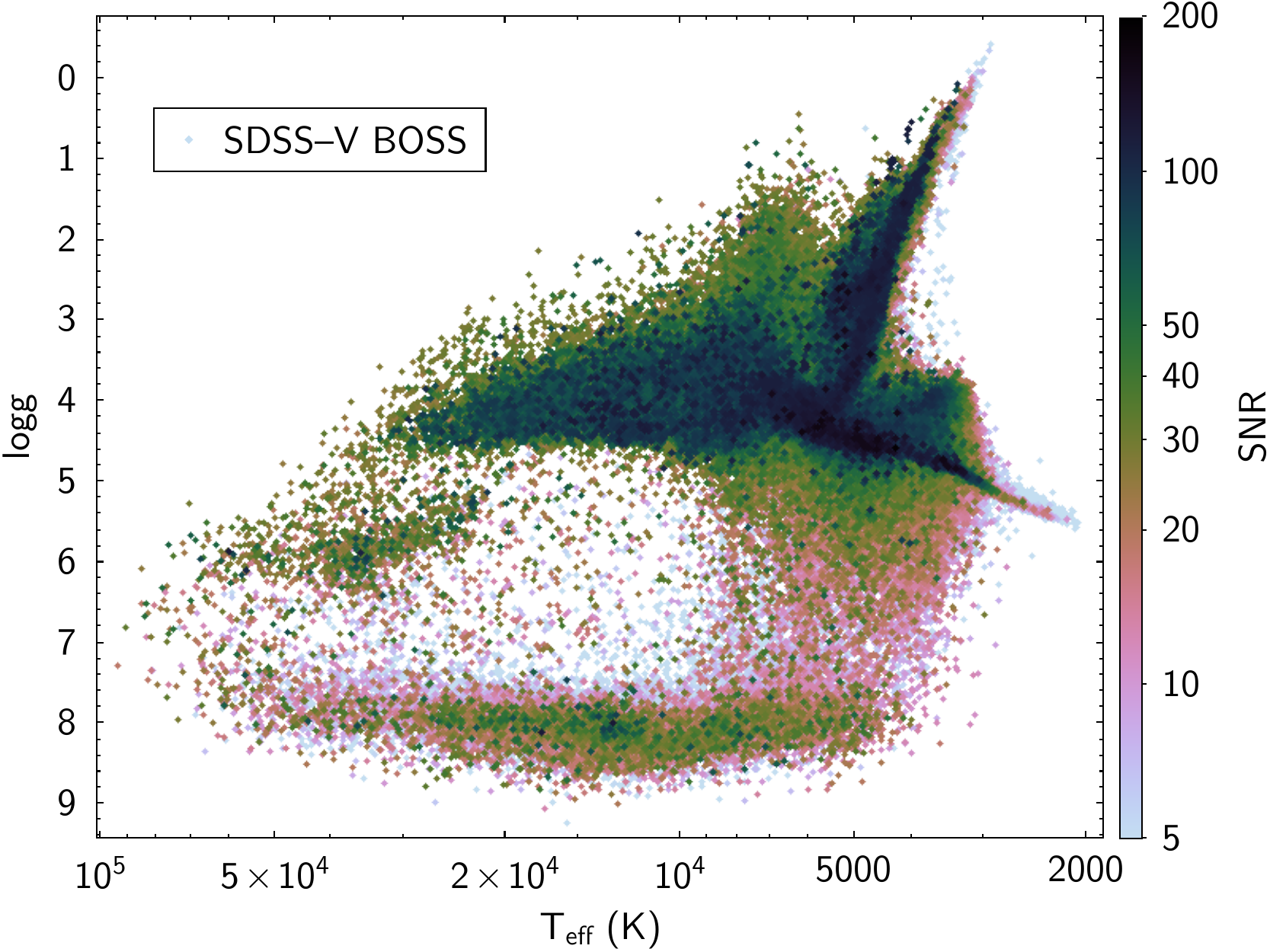}
\plotone{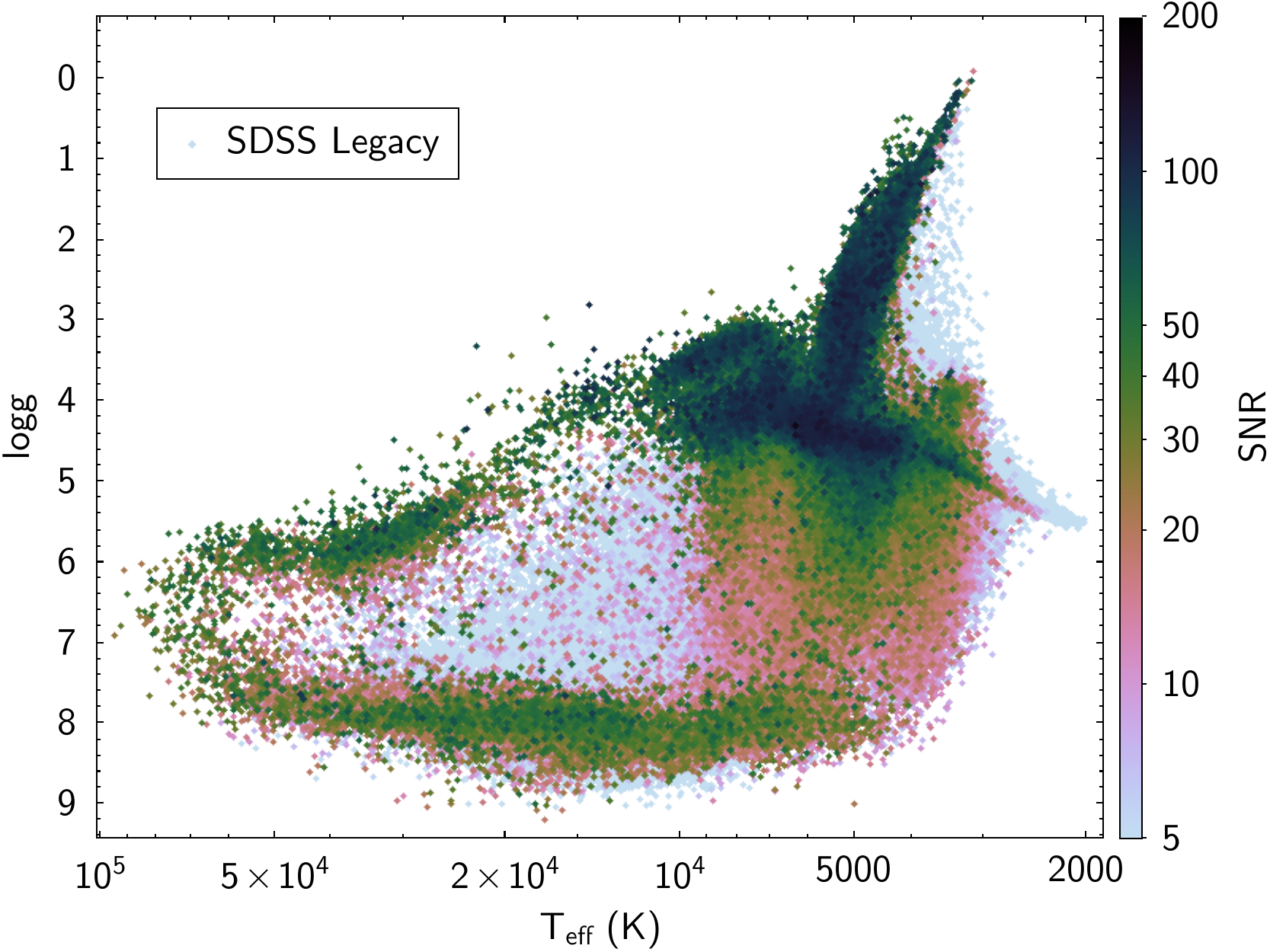}
\plotone{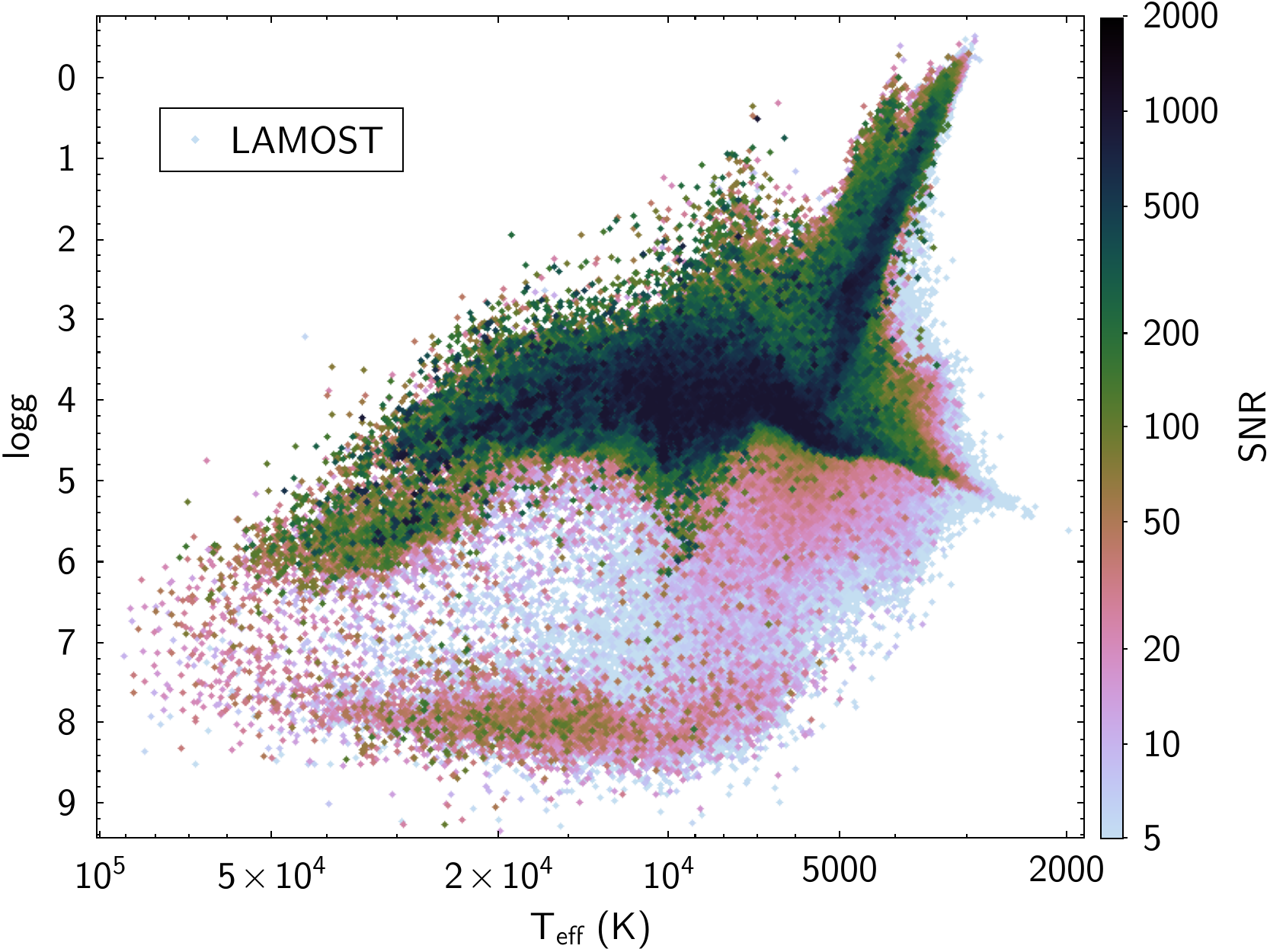}
\caption{The distribution of the derived \teff\ \& \logg\ in the BOSS spectra, color coded by SNR.
\label{fig:snr}}
\end{figure}

We apply BOSS Net to all of the stellar spectra observed by BOSS to date. This does include the data originally used for training, but, as BOSS obtains spectra of several thousand stars on a nightly basis, the majority of the data are new.

We derive the uncertainties in the predictions following a similar implementation to the previous iterations of the APOGEE Net. In particular, we generate 20 different realization of the same spectrum by scattering the input fluxes by the reported uncertainties. To ensure the stability of the model, if uncertainties are larger than 5 times the median across the spectrum (such as, e.g., in the regions dominated by the telluric lines, or near the edges of the spectrum), they are capped to that level. All these realizations are separately passed through the model. The scatter in the resulting predictions is then evaluated and adopted as the uncertainties. These uncertainties are model-dependent and are not representative of systematic errors which can be assessed through comparison to external datasets. However, the reported uncertainties, do provide meaningful variance.

The example spectra plotted as a function of the predicted \teff\ are shown in Figure \ref{fig:boss_teff_comp}. The resulting distribution of \teff\ and \logg\ is shown in Figure \ref{fig:snr}. It provides a good coverage of the underlying parameter space that is present in the training data. There is some degree of scatter in \logg\ of cool stars with extremely low SNR being stranded in the parameter space between the main and the white dwarf sequences; these sources can be identified through \logg\ uncertainties.

Figure \ref{fig:snr} also highlights the difference in the selection function between SDSS-V BOSS observations, those conducted in prior SDSS iterations, and those done with LAMOST. LAMOST does not go to as faint magnitudes as BOSS, lacking very cool brown and white dwarfs. In large part, it is also less targeted towards the more exotic objects, while, e.g., pre-main sequence stars or some of the largest blue supergiants are present in LAMOST data due to the sheer number of source it observed, they compose a significantly smaller fraction of the catalog.

On the other hand, legacy SDSS optical observations were primarily motivated by extragalactic sources, they are pointed high above the galactic plane. They are preferentially more metal poor, with almost half the sample with [Fe/H]$<$-1. They lack high mass stars, since they are unlikely to migrate far out of the disk due to their short lifetime. Moreover, being a deep survey, high mass stars would also be preferentially excluded due to their brightness. There is also significant representation from compact objects. In particular, subdwarfs can be clearly be seen as a continuation of the horizontal branch, which is consistent with interpretation of their origin \citep{heber2008}.

\begin{figure*}
\epsscale{1.1}
\plottwo{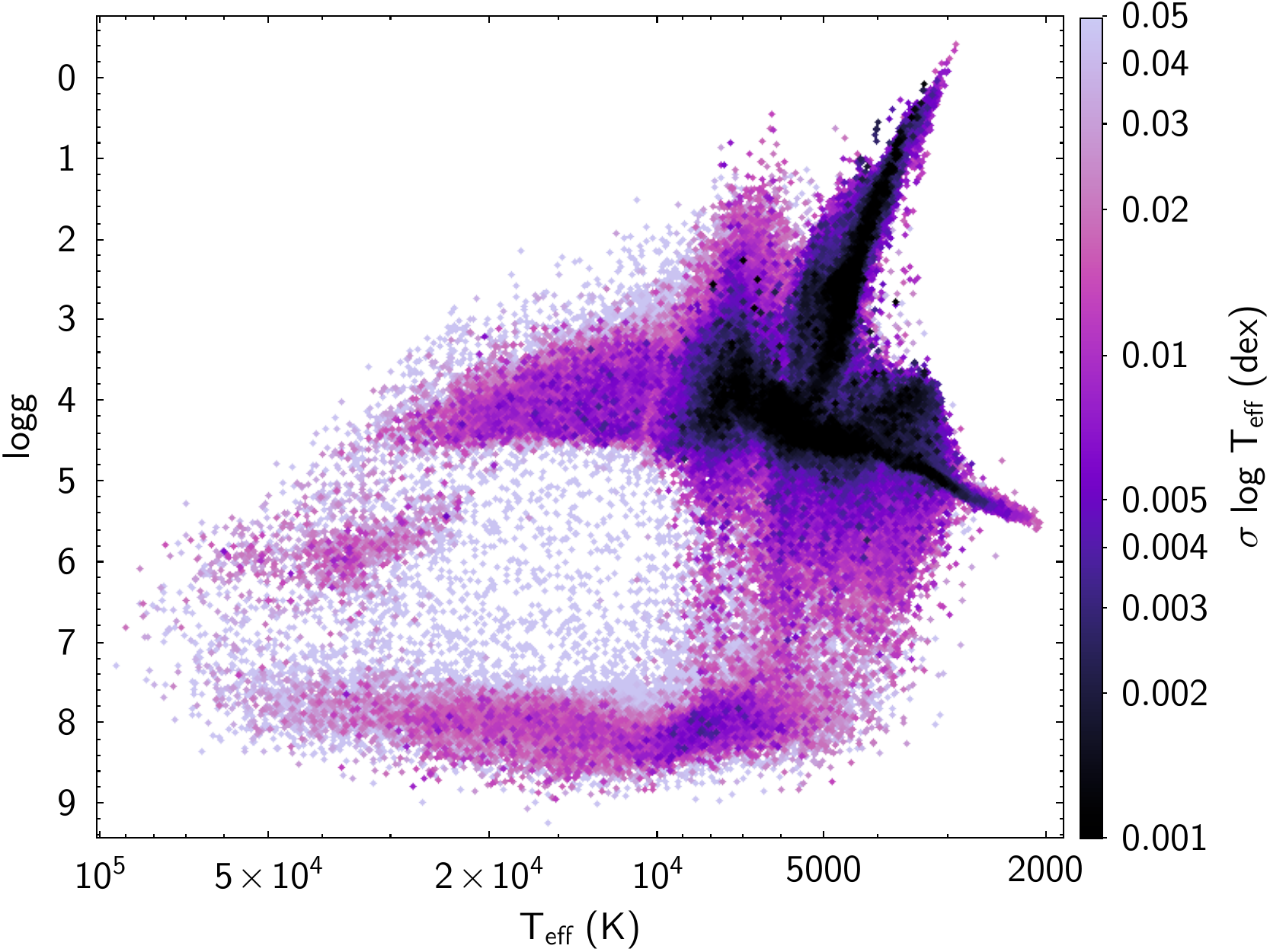}{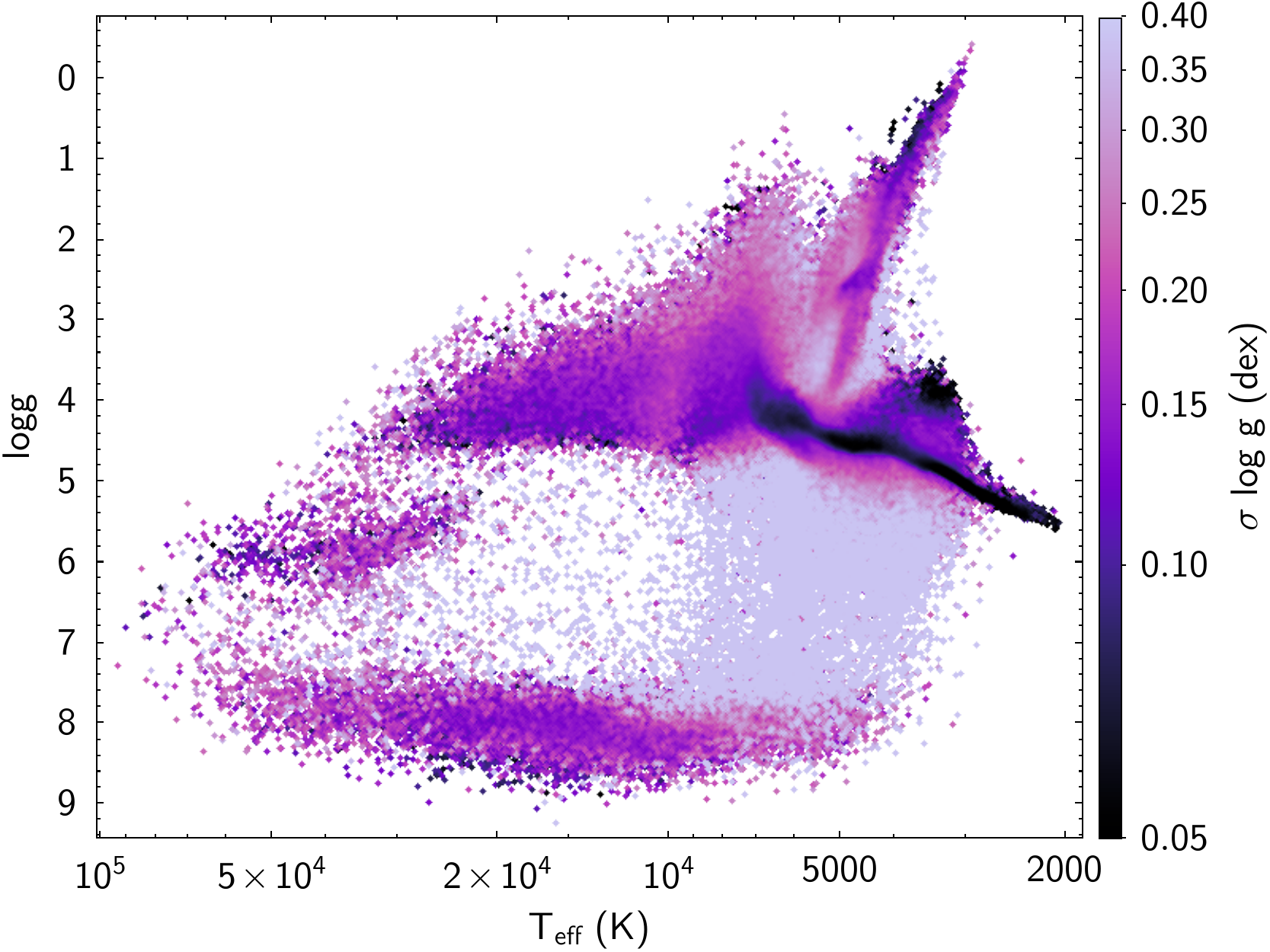}
\plottwo{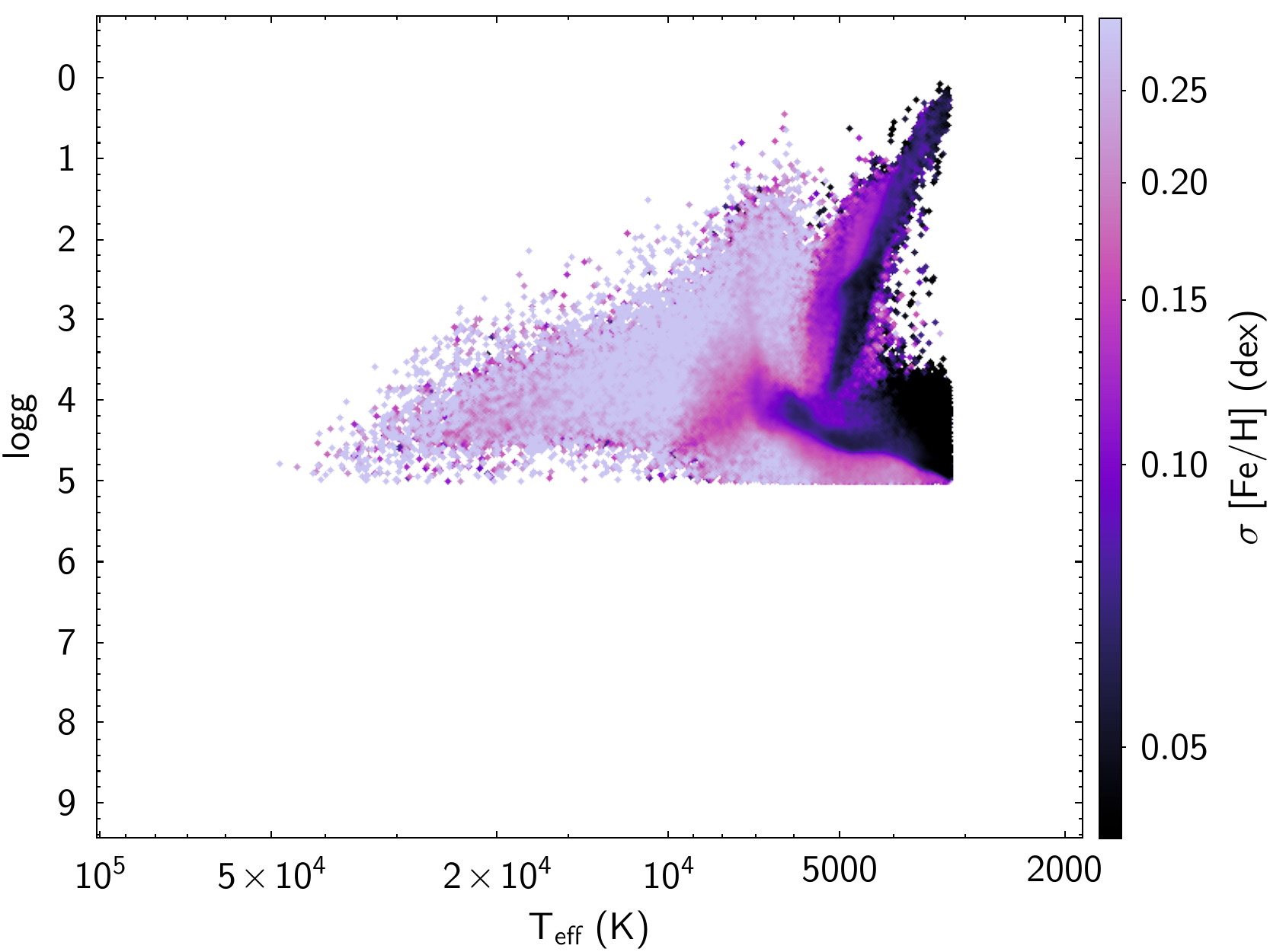}{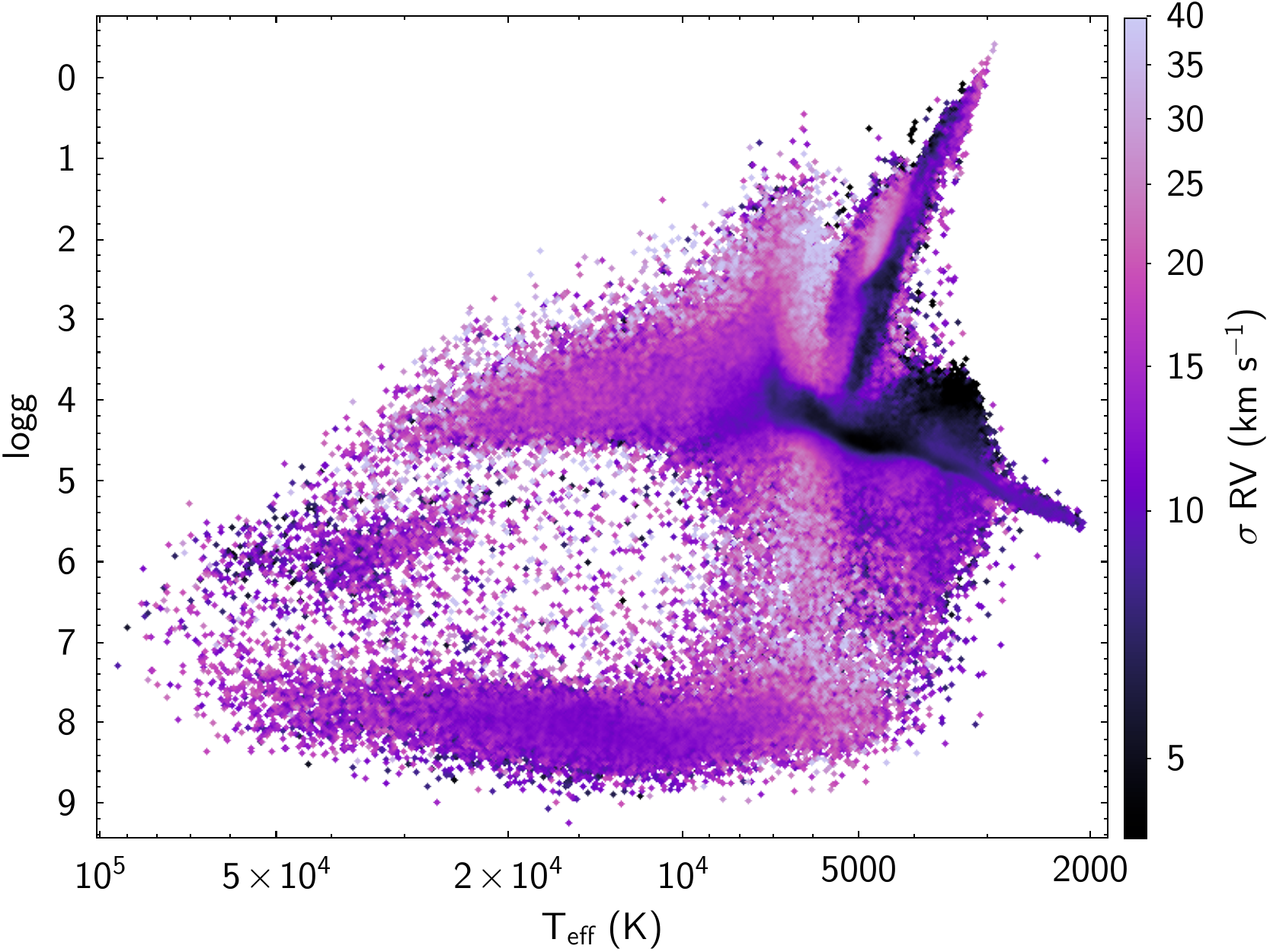}
\caption{Typical uncertainties in \teff, \logg, [Fe/H], and RV across the full parameter space.
\label{fig:sigma1}}
\end{figure*}

The typical uncertainties in the \logg\ for the sources with SNR$>15$ are 0.09 dex for the cool stars (\teff$<$6700 K) and 0.13 dex for hot stars. In \teff\ they are typically 0.007 and 0.02 dex in cool and hot stars. In [Fe/H] they are 0.07 and 0.16 dex, respectively. In RV they are 7 and 12.5 \kms (Figure \ref{fig:sigma1}).

\begin{figure}
\epsscale{1.2}
\plotone{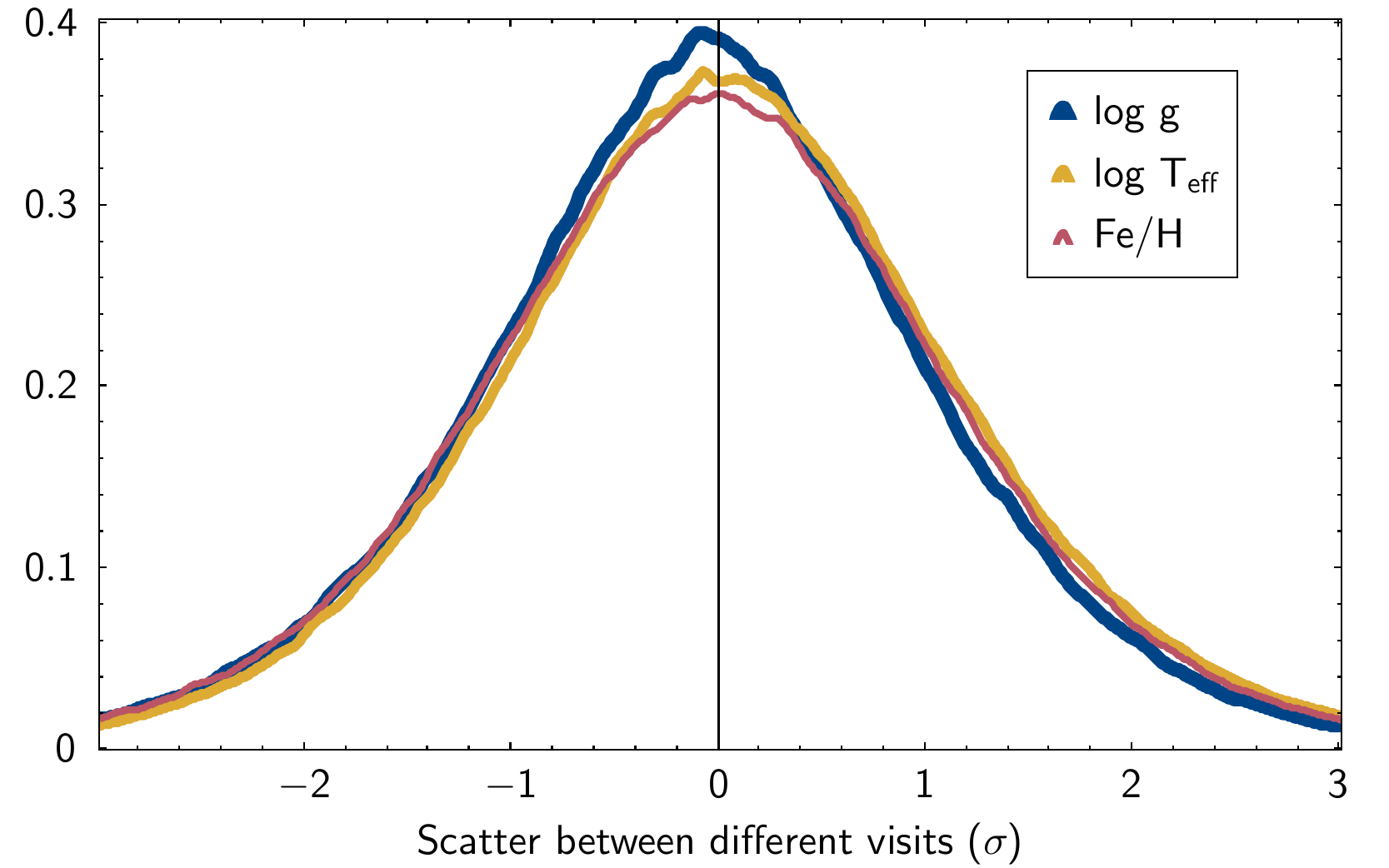}
\caption{Difference between the predictions obtained for the same stars with multiple visits, divided by the uncertainties in the predictions (added in quadrature). Blue curve shows the resulting distribution of the scatter in \logg, yellow in \teff, and red in [Fe/H].
\label{fig:sigma}}
\end{figure}

Several sources have been observed multiple times, enabling testing the consistency of the predictions and the accuracy of the resulting uncertainties (Figure \ref{fig:sigma}). In general, the resulting scatter is well replicated by the reported errors, with the full width of half maximum (FWHM) being almost exactly 1$\sigma$. Comparing parameters in the sources observed by both BOSS \& LAMOST, the scatter is somewhat larger, typically 1.2-1.5$\sigma$, but, nonetheless, both set of parameters are very consistent with one another (Figure \ref{fig:test1}).

In comparing SDSS-V and LAMOST RVs to other surveys such as APOGEE, BOSS Net appears to be significantly more stable in low SNR regime than the native data reduction pipelines for these instrument, nor does there  appear to be any position of the sky dependent systematic offsets as those that were originally apparent in LAMOST data. They do however remain in the legacy SDSS I-IV BOSS spectra, with typical offsets on the order of $\sim$5 \kms. Despite being interpolated onto a common grid, BOSS Net was able to learn how to differentiate between BOSS and LAMOST spectra, in order to correct for LAMOST offsets. But, since SDSS I-IV spectra use the same instrument as SDSS-V, the inconsistency in the data reduction strategy has resulted in an inconsistent calibration of RVs.

In comparing the parameters produced by BOSS Net to other pipelines developed within SDSS-V, we generally achieve good agreement in most cases. In some cases, e.g., RVs of white dwarfs, BOSS Net solutions have higher uncertainties than what is possible to achieve through using more specialized pipelines. Additionally, since BOSS Net does not normalize the continuum prior to determining parameters, caution should be exercised with early type stars that are very extinct (\teff$>$7000 K \& $A_G>3$). Extinction is not as significant of a concern for later type stars, since they are very rich in spectral features, moreover, with high degree of extinction they are unlikely to be observed within the magnitude limit.

\section{Discussion} \label{sec:discussion}

\begin{figure*}
\epsscale{1.0}
		\gridline{\fig{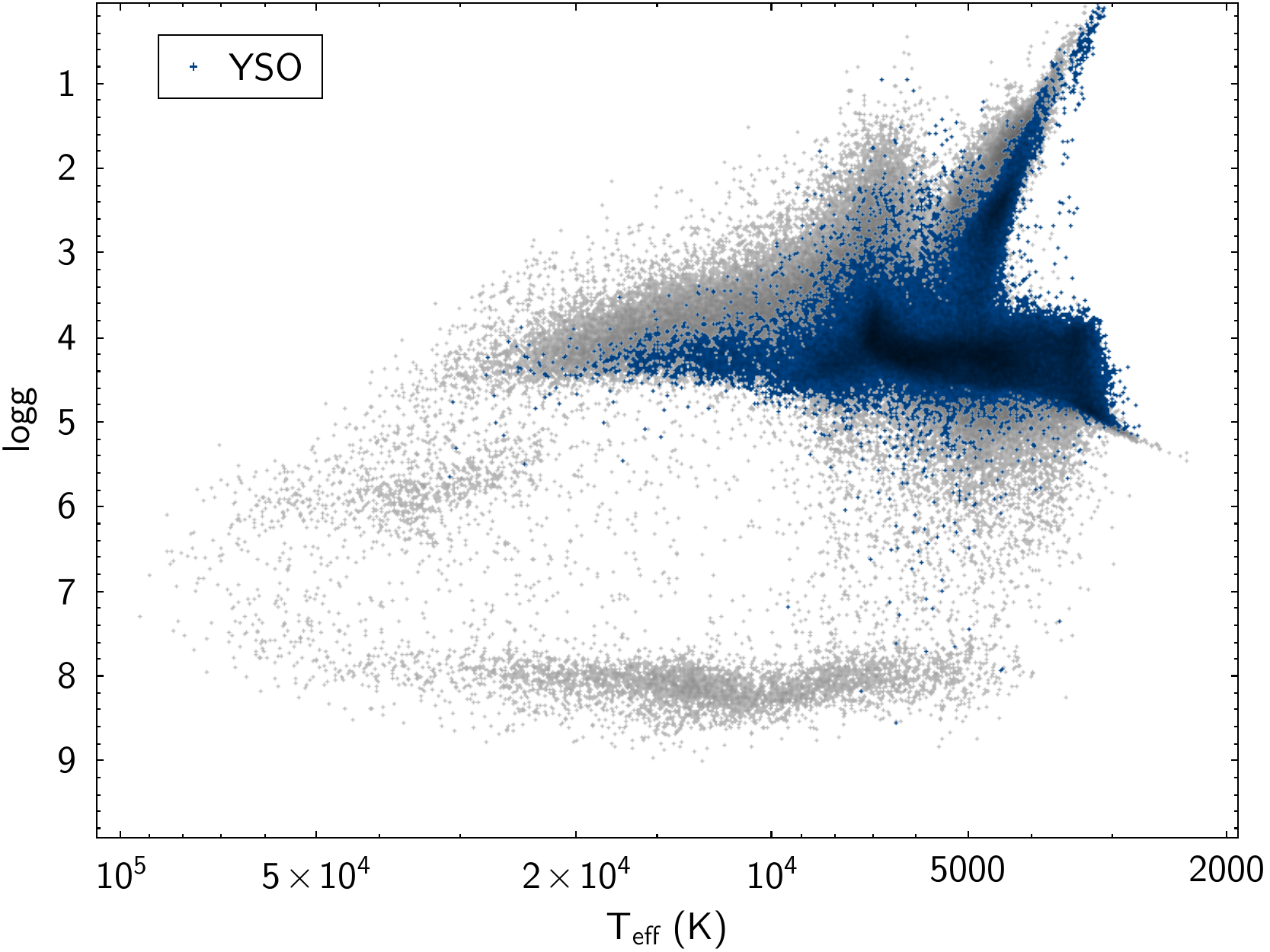}{0.5\textwidth}{}
		          \fig{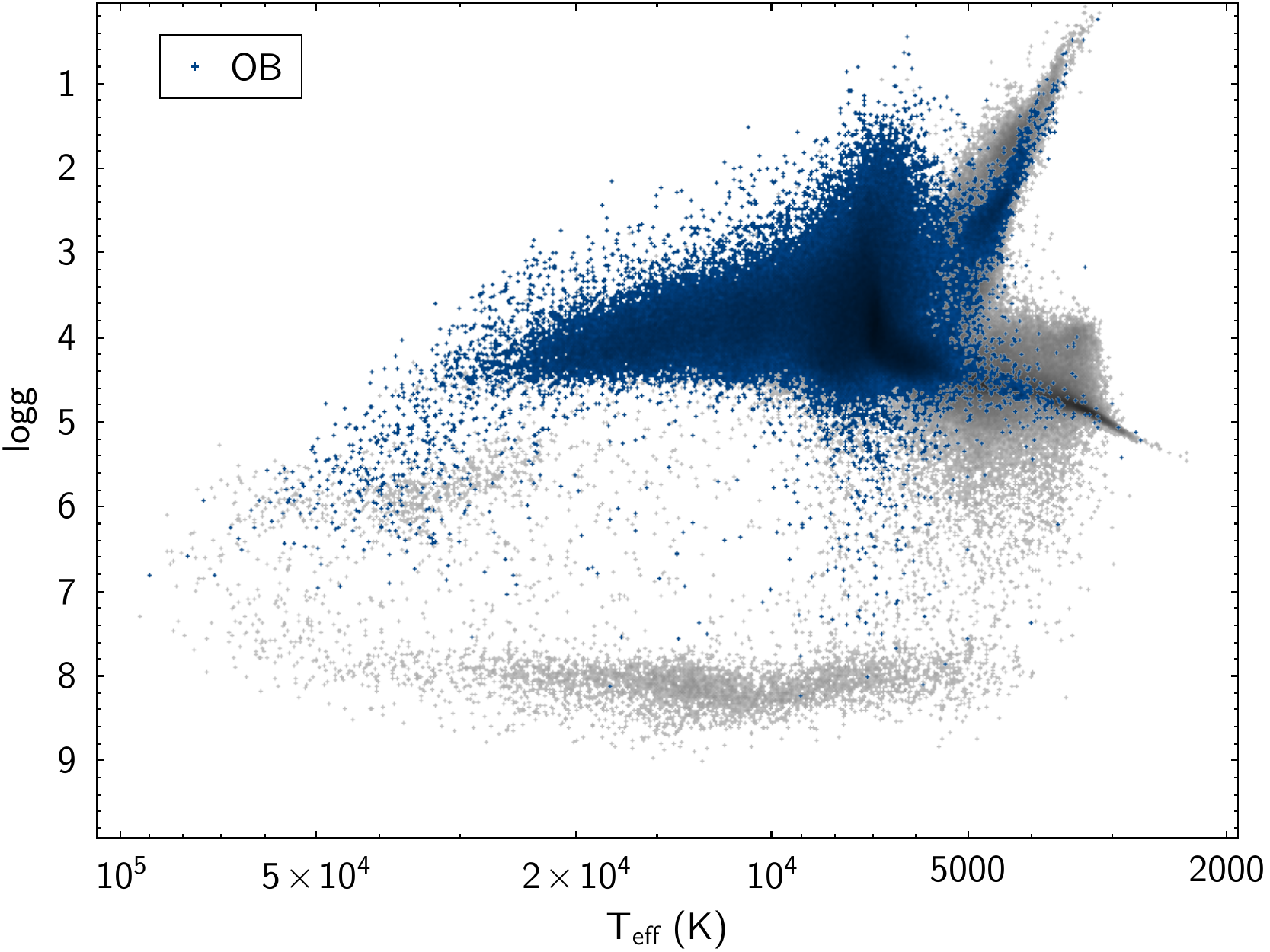}{0.5\textwidth}{}
        }\vspace{-0.8cm}
        		\gridline{\fig{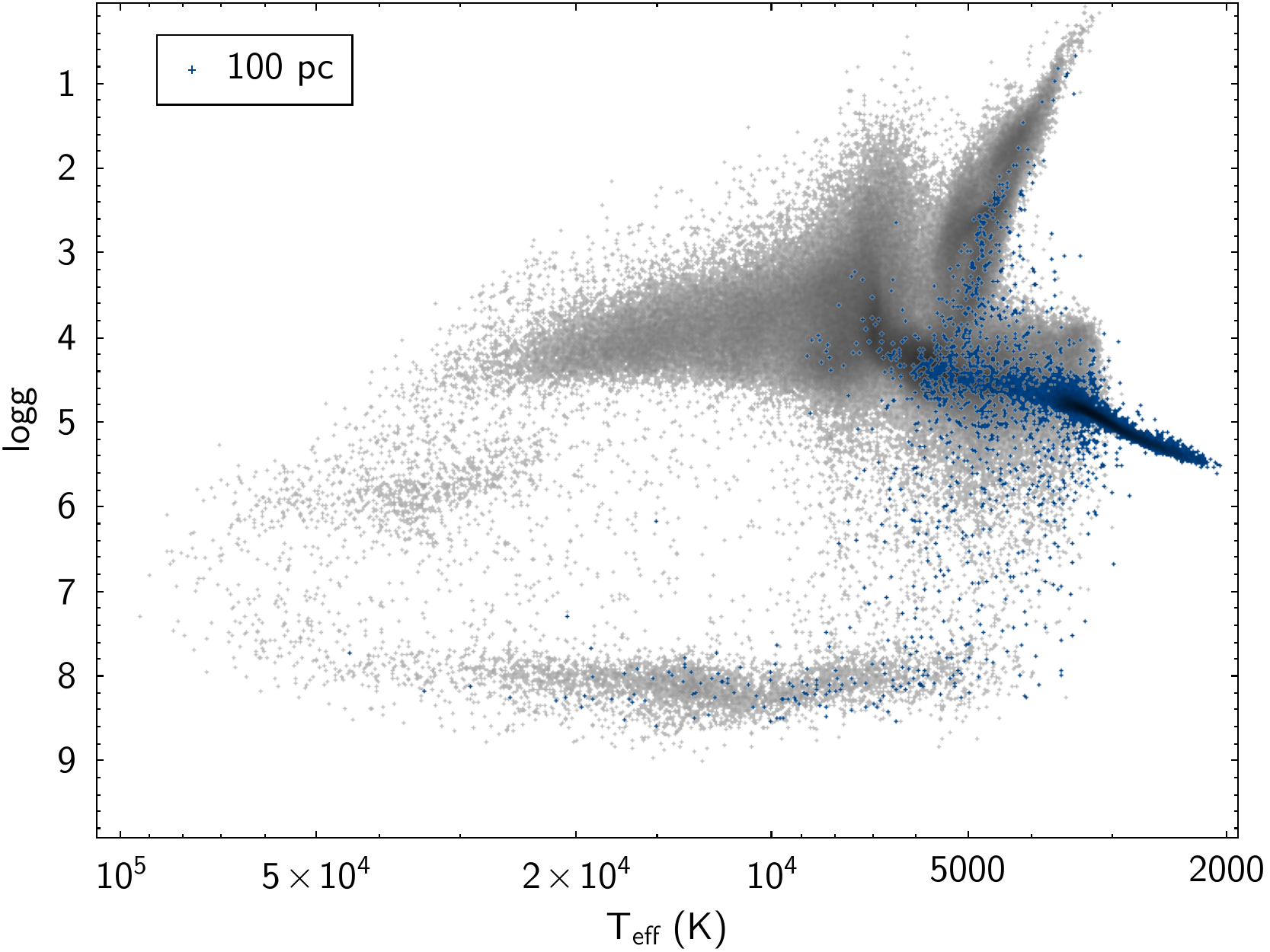}{0.5\textwidth}{}
		          \fig{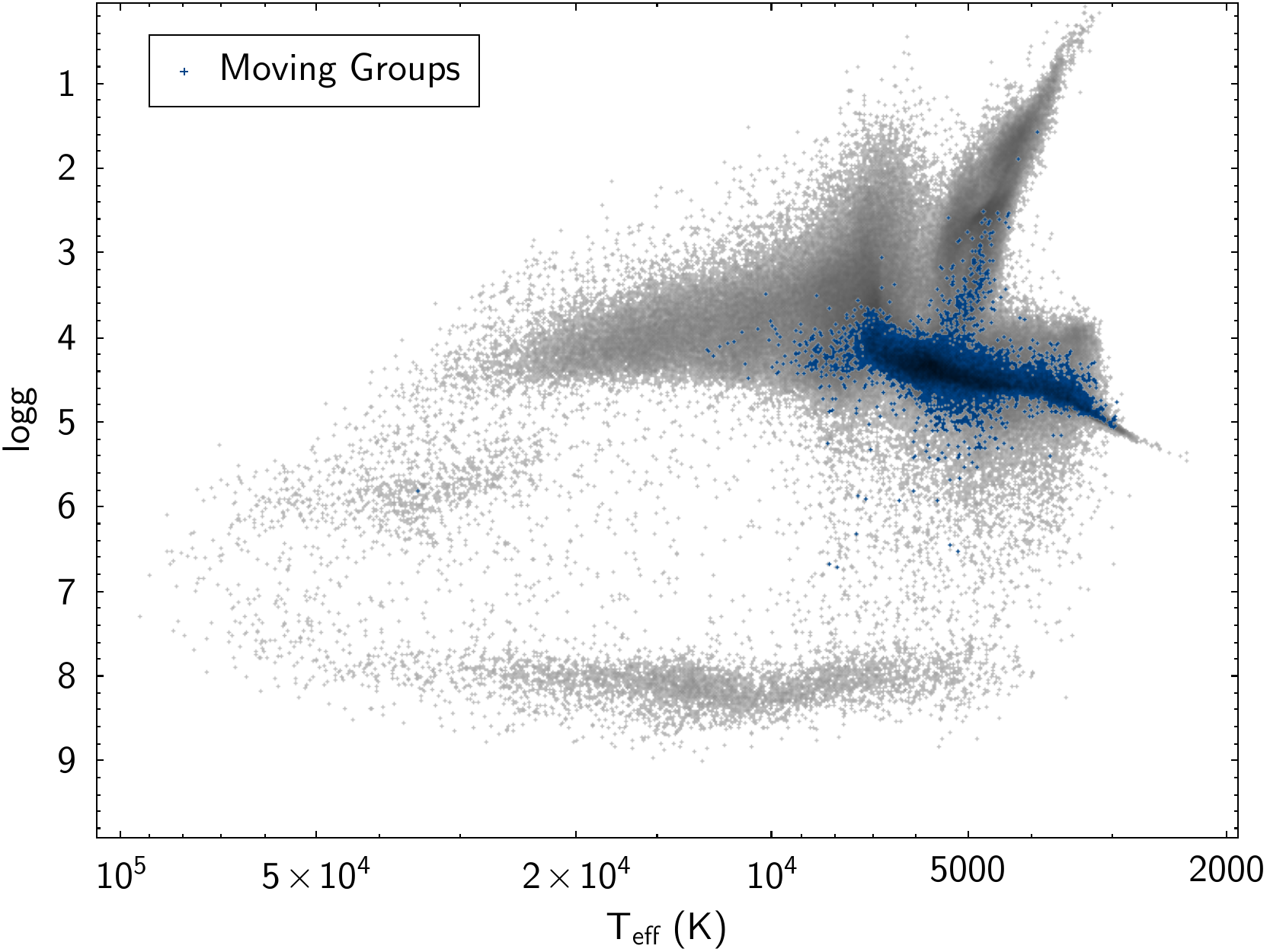}{0.5\textwidth}{}
        }\vspace{-0.8cm}
        		\gridline{\fig{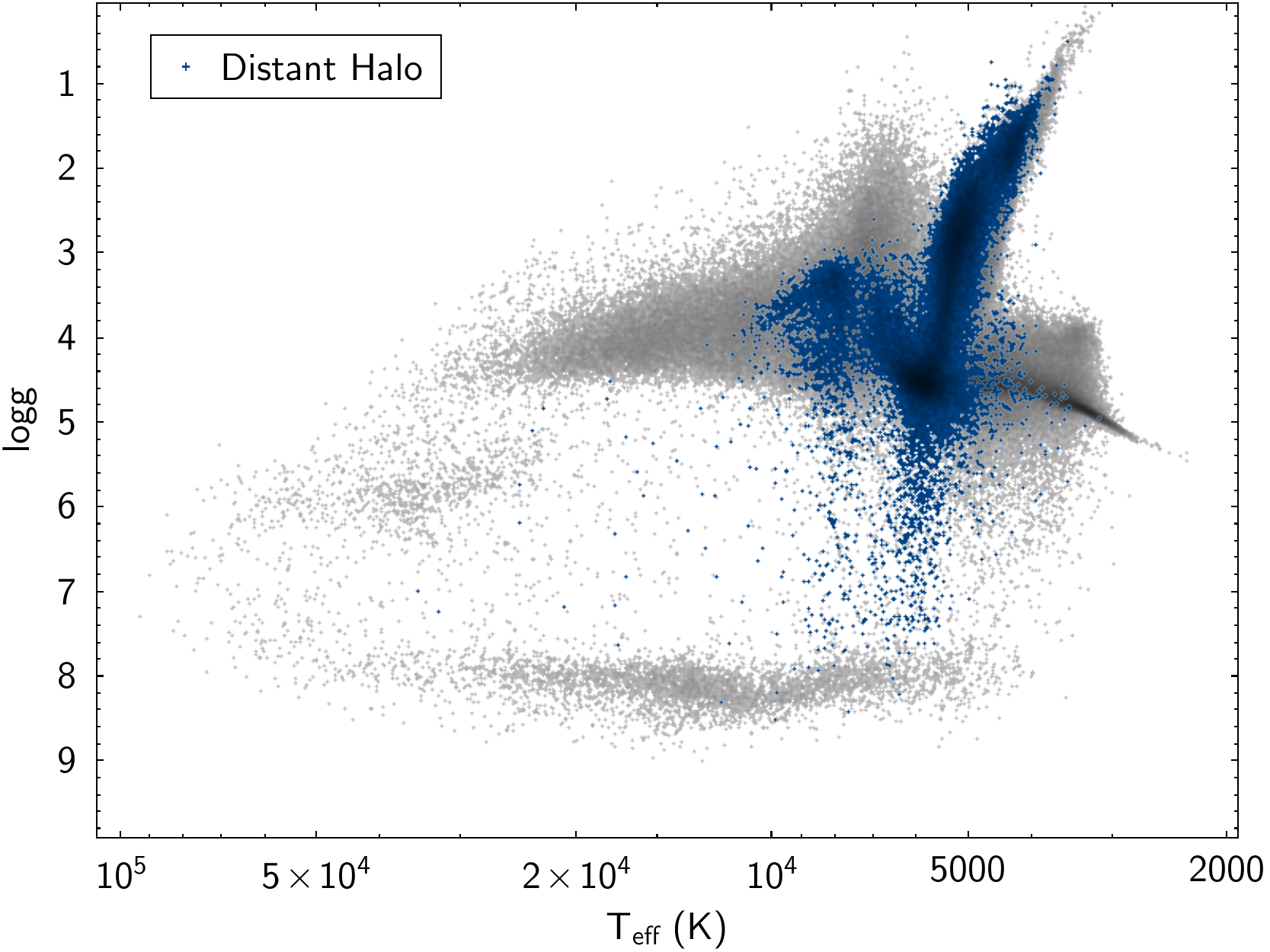}{0.5\textwidth}{}
		          \fig{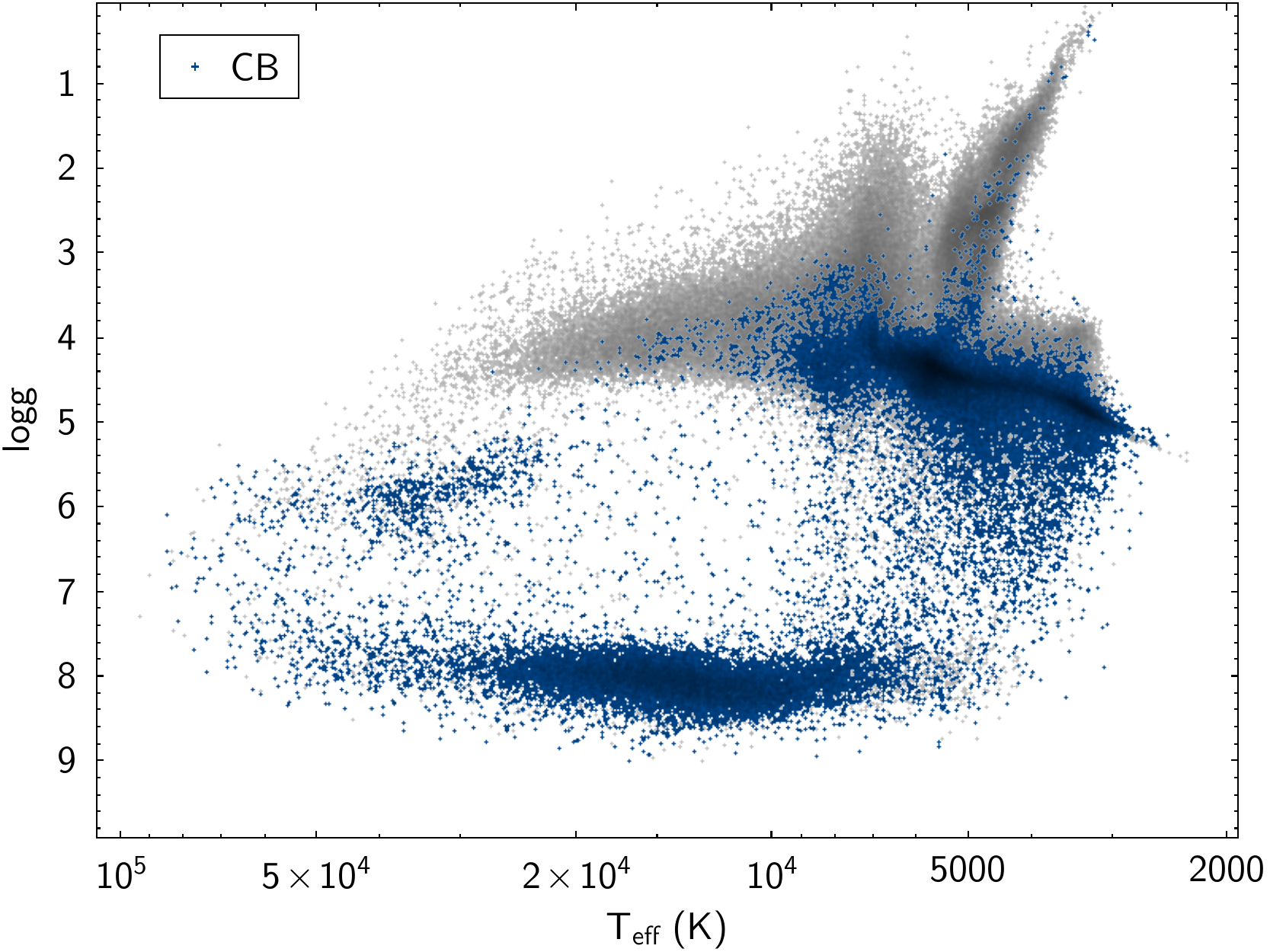}{0.5\textwidth}{}
        }\vspace{-0.8cm}
%        		\gridline{\fig{grid_legacy.pdf}{0.5\textwidth}{}
%		          \fig{grid_f.pdf}{0.5\textwidth}{}
%        }\vspace{-1cm}
\caption{The distribution of \teff\ and \logg\ of sources targeted by various SDSS-V programs, including young stars ($<$30 Myr), OB stars, sources within 100 pc, members of moving groups ($>$30 Myr), sources in the halo with distance $>$2 kpc (including K giants, horizontal branch stars, RR Lyr, and metal poor dwarfs and giants), and compact binaries.
\label{fig:grid}}
\end{figure*}

SDSS-V consists of several dedicated programs, each focused on different types of stars (Figure \ref{fig:grid}), such as, e.g., nearby dwarfs, young stars, compact binaries, distant red giants, OB stars, and many others \citep{almeida2023}. Each program has been responsible for producing a catalog of likely candidates that represent their sources of interest; these catalogs are used in targeting for the survey. As such, every spectrum observed by SDSS-V has a flag specifying which program it has been targeted by, separating objects into different classes. Of course, this preliminary tagging is imperfect and prone to contamination, but, examining the derived spectroscopic properties of the sources grouped by their targeted program in bulk enables evaluating the performance of BOSS Net for different types of stars. Here we examine sources from just a few programs that are representative of the underlying parameter space.

\begin{figure}
\epsscale{1.1}
\plotone{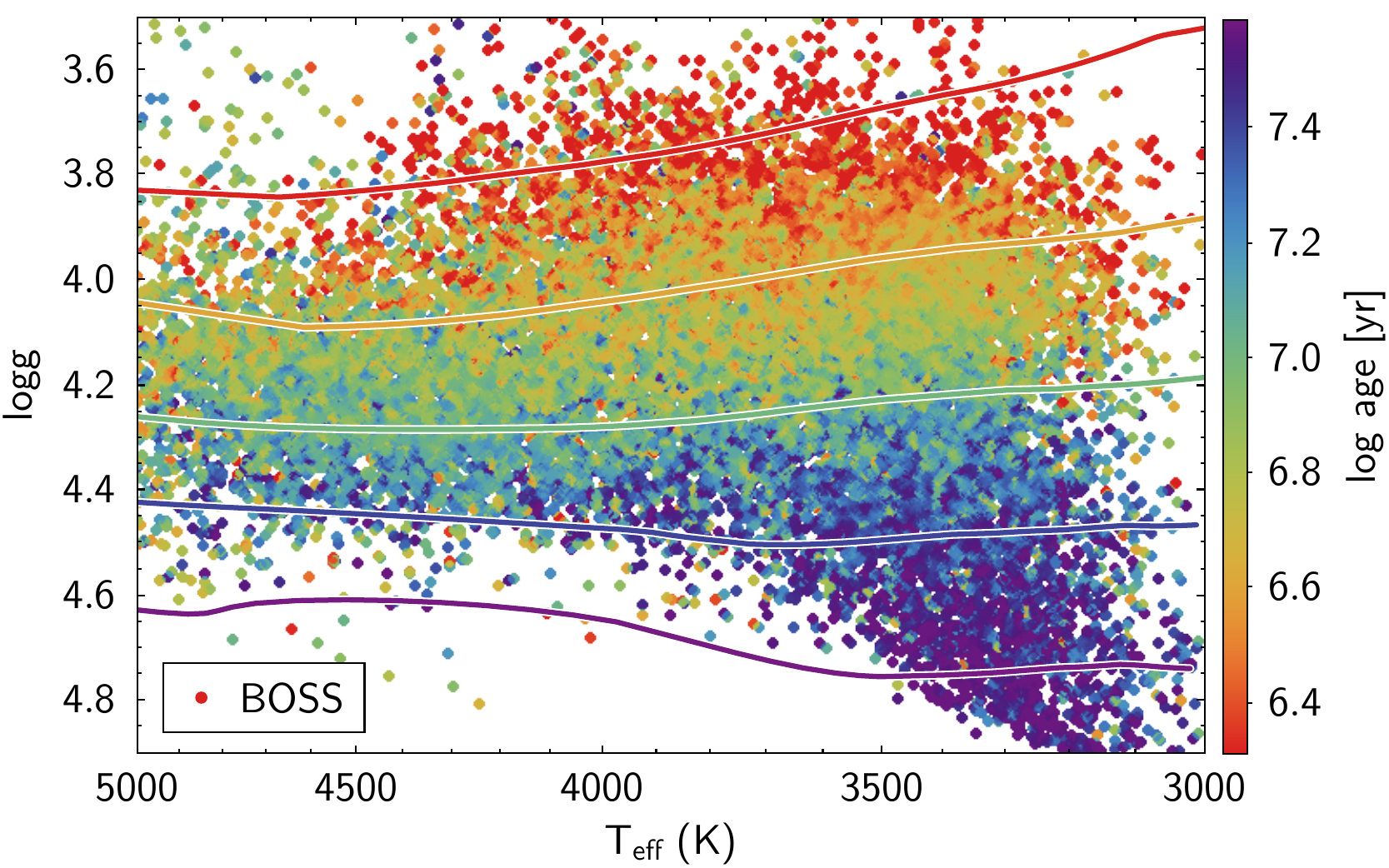}
\plotone{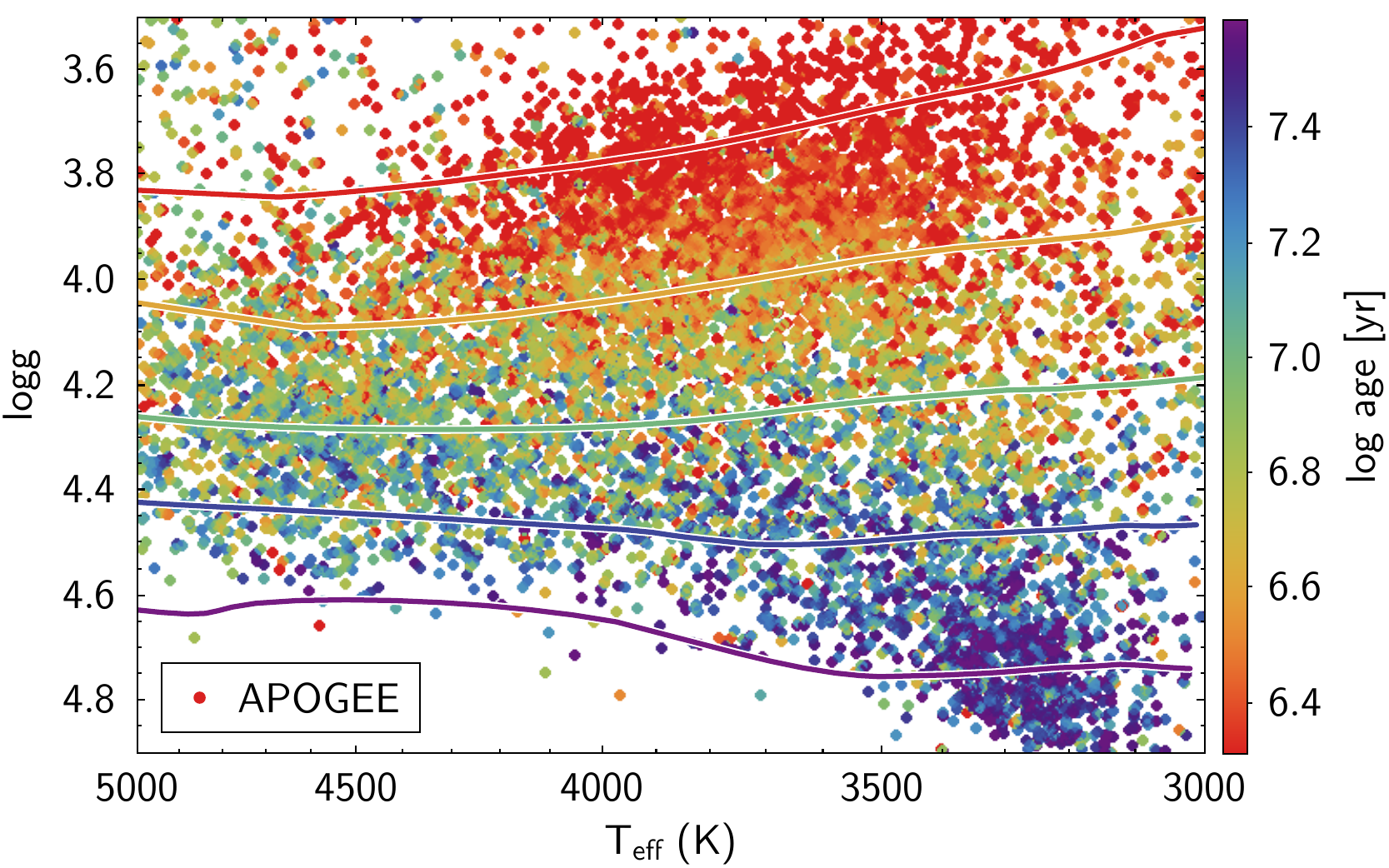}
\caption{The distribution of \teff\ and \logg\ of the pre-main sequence stars, color coded by photometrically derived ages using Sagitta \citep{mcbride2021}. The lines with white outlines show MIST isochrones \citep{choi2016} with ages of 6.2, 6.6, 7.0, 7.4, and 7.8 dex (from top to bottom). Note that \logg s independently show a strong correlation with age, and that the spectroscopic parameters are consistent with the photometric estimates. Top panel: outputs from BOSS Net. Bottom panel: outputs from APOGEE Net (Appendix \ref{sec:apogee}); the apparent difference on the youngest end is driven primarily by the selection function of the sources observed in each wavelength regime to-date.
\label{fig:pms}}
\end{figure}

\begin{figure}[!t]
\epsscale{1.1}
\plotone{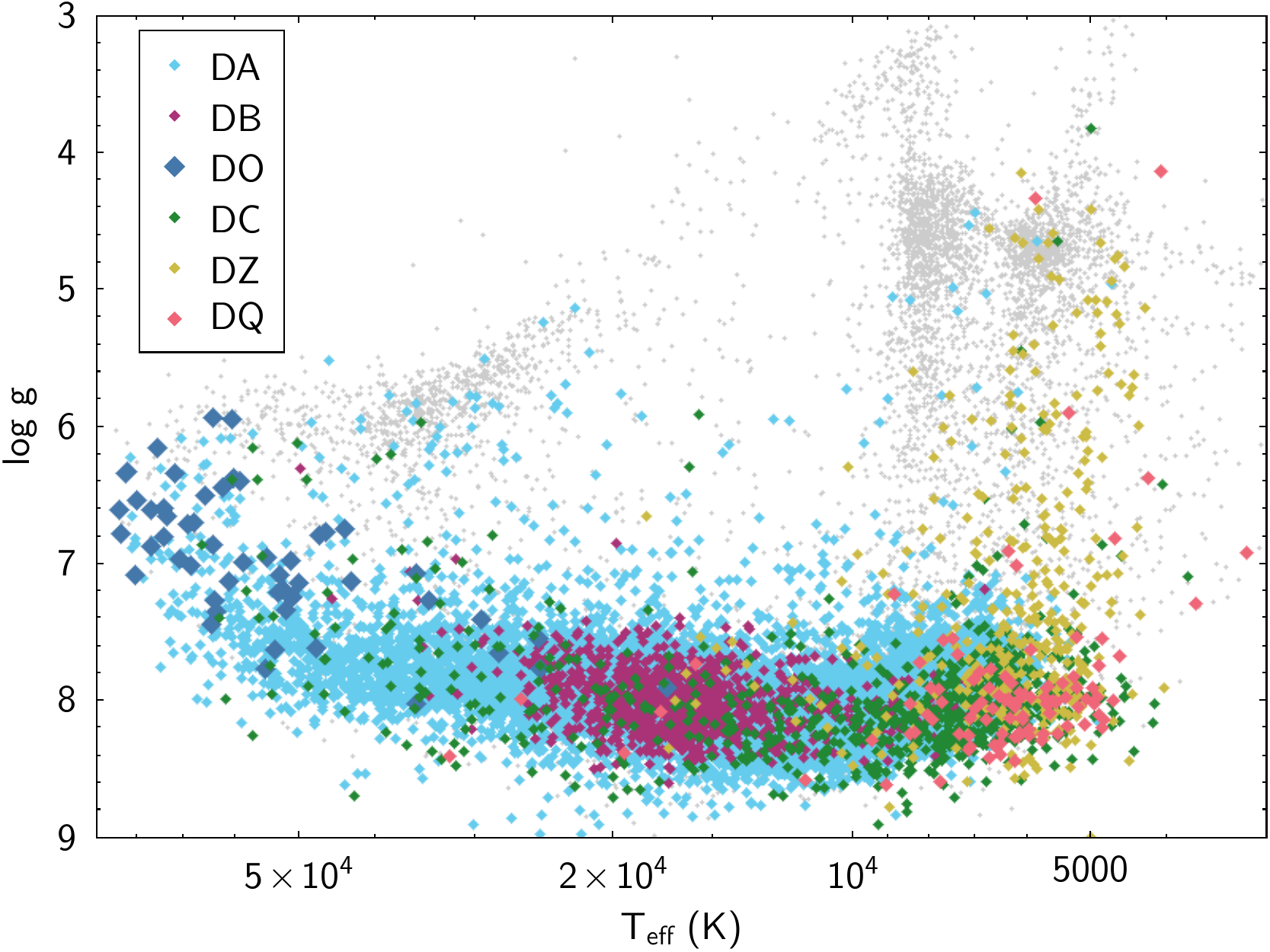}
\caption{Derived spectroscopic parameter space of white dwarfs in legacy SDSS I-IV data, color coded by the source classification from \citet{gentile-fusillo2021}.
\label{fig:wd}}
\end{figure}

The stars observed by the young star program \citep{kounkel2023} are particularly useful in testing \logg. The age of these stars (typically $<$30 Myr) can be estimated through photometry alone, be it through isochrone fitting, or through a data-driven approach \citep[]{mcbride2021}. As such, given the rapid evolution of the pre-main sequence low mass stars, their parameters can be evaluated from the comparison to the theoretical models of the stars of that age. Indeed, spectroscopically derived \logg\ values presented here demonstrate an excellent agreement with the MIST isochrones \citep[Figure \ref{fig:pms}]{choi2016}. The calibration here is significantly improved, and does not appear to show any systematic trends relative to the models that were present in the previous iterations of APOGEE Net \citep{olney2020}.

High mass stars quickly settle onto the main sequence, thus it is difficult to observe them in a pre-main sequence phase. However, comparing the position of the young stars to those targeted by the OB star program \citep{zari2021} shows that the former tend to be found at higher \logg\ values (Figure \ref{fig:grid}). High mass stars have short lifetimes, younger sources will be found on the main sequence, but they will evolve away from it in as little as $<$100 Myr. As they transition to giants, they will become more luminous, as such more likely to be targeted. As such, it is unsurprising that the stars within the OB program tend to be predominantly more evolved than those in the young star program.

Sources observed by the solar neighborhood census program (targeting sources within 100 pc based on Gaia parallaxes) are predominantly M dwarfs and brown dwarfs. Earlier type stars are unlikely to have been selected, as, due to their proximity, they are brighter than the (current) bright limits of BOSS, which would result in saturation. On the other hand, other programs are unlikely to contain brown dwarfs, since at larger distances they become increasingly too faint.

Some observations were conducted of members of various moving groups from \citet{kounkel2020}. These populations are found within 3 kpc, with ages ranging between 30 Myr and 4 Gyr. The observed sources are predominantly found on the low mass part of the main sequence, since they are old enough to have settled there, but still sufficiently young to have not evolved along the red giant branch. Similarly to the above, hotter stars tend to be too bright for BOSS to currently observe. The resulting sequence is relatively tightly concentrated in \logg, with some scatter predominantly from the [Fe/H] spread. To date, there are only a few groups with spectroscopic data for more than a handful sources. With a growing census in the future, it would be possible to use them to evaluate the stability in [Fe/H].

Stars in the halo have been targeted though several different methods, here we focus on just a few of them, restricting the sample to distant stars with distances $>$2 kpc. Sources that have been identified as metal poor dwarfs are typically found with higher \logg\ than the more metal rich main sequence stars of comparable \teff\ in other programs. Similarly, metal poor giants are hotter than the metal rich giants. Both of these are expected given evolutionary models \citep[e.g.,][]{choi2016}, and the resulting [Fe/H] also reflect this difference.

Compact binaries have been typically selected based on their UV excess, to preferentially select systems containing white dwarfs. A sizeable fraction of these sources have \logg\ consistent with being main sequence stars, since this would typically be a significantly brighter star the system. However, sources where the spectrum is dominated by the flux of white dwarfs or hot sub-dwarfs are also easily apparent in the sample.

We similarly compare the parameters for the legacy SDSS spectra for white dwarfs to the classification from \citet{gentile-fusillo2021} in Figure \ref{fig:wd}. We well recover the expected temperature differences between different classes. Almost all of the sources in that catalog that are confirmed to be WDs (as opposed to stars, hot subdwarfs, and other types of objects) are indeed distinguishable based on their \logg, although DZ (in contrast to similarly cool DQ) type WDs may be most succeptible to being caught in between WD and main sequences.

Most of these trends are not unexpected, and indeed, many of these types of sources were originally present in the training set, even though in some cases particular class of sources might have been split across different catalogues. Nonetheless, the volume of the underlying dataset and the self-consistency of the derived labels does make the comparison between the sources to be more illuminating.

%Furthermore, through interpolating across the sources of all types, the model was able to independently identify differences in the parameter space of certain type of sources that it was not familiar with. For example, although there are several different types of white dwarfs, to date, only DA-type white dwarfs had a large quantity of objects for which \teff\ and \logg\ have been previously measured. DB type sources were not present in the training set. Nonetheless, the trained BOSS Net model was able to separate DA and DB sources. At 30,000 K, DB sources have \logg\ between DA white dwarfs and hot subdwarfs, though a the cooler \teff\, the two sequences merge (Figure \ref{fig:wd}).

\section{Conclusions} \label{sec:conclusions}

We present BOSS Net, a model for evaluating spectroscopic stellar parameter from optical spectra, namely, from BOSS and LAMOST. This model is capable of deriving \teff, \logg, and [Fe/H] across all of the stellar objects observed by these instruments in the range of 1700$<$\teff$<$100,000 K and 0$<$\logg$<$10 in a self-consistent manner. This includes main sequence stars (from OB stars to brown dwarfs), pre-main sequence stars, evolved stars (both hot and red giants), hot subdwarfs, as well as white dwarfs. In pre-main sequence stars, the resulting parameters are calibrated to the stellar evolutionary models, and \logg\ can be used as an independent indicator of the age of the star.

This model has been built using the training set consisting of carefully assembled catalogues produced by a number of different studies that specialize in specific type of sources. The narrow focus of these studies has provided crucial expertise in characterizing the specific features in the parameter space of these stars. For example, comparing the spectra to the synthetic templates makes it possible to understand the fundamental physics behind various stellar objects, however, such an approach has its limitations. Each one has a boundary in the parameter space it explores, and joining the resulting parameters can be non-trivial without introducing systematics that complicate the interpretation of these parameters near the boundaries. Synthetic spectra may also miss vital spectroscopic features that are present in the real data that further add systematics.

The data driven approach presented here enables us to bridge across all of these studies, allowing the model to improve on the self-consistency of the input parameters. Stars of different types are still governed by the same physics, even though they occupy different regions of the parameter space, their inclusion makes the model more general. The ability to characterize all of the stellar spectra within a single model is vital in large surveys, as this enables the model to take advantage of all available data. Data-driven models also allow the model to translate the parameter space between different surveys and different instruments. Furthermore they are fast and efficient, which is also an important consideration for data processing in a large survey.

BOSS Net will be available publicly on GitHub, and it is also being incorporated into Astra, which is the analysis framework for the Milky Way Mapper within SDSS-V, and which manages various data processing pipelines. As a result, the parameters that it produces will be made available in the subsequent data releases.

There are a number of more specialized pipelines within Astra, including pipelines focused on low mass stars, white dwarfs, hot stars, metal poor stars, etc. At the time of the data releases, an assessment by the working groups within the survey will be made regarding the recommended set of parameters for each stellar class.

We do note that while this pipeline was designed to be capable of processing all of the stellar spectra in SDSS to provide a homogeneous set of stellar parameters, it was primarily developed to support pre-main sequence stars, and it may not be a one-size-fits-all for all applications. It may be unreliable in hot stars that are highly extinct ($A_G>3$ \& \teff$>$7000 K). The reported parameter space for the cool white dwarfs likely has significant systematics, thus, using stellar parameters from specialized pipelines would be recommended for these sources. Reasonable caution should also be exercised with regards to the more exotic types of stars that have not been accounted for in the training set: if they overlap with the previously explored parameter space, there may be systematic offsets in their absolute calibration. If they are outside of the parameter space covered by the training set, the predicted parameters could become unphysical, or they could be aliased to the stars with most similar features. In such cases a comparison with an independently derived set of parameters from different pipelines may be advised. In future it may be possible to isolate these sources to a separate catalog.

\begin{appendix}

\section{Overview of the training set}\label{sec:overview}
\begin{figure*}
\epsscale{1.1}
\plottwo{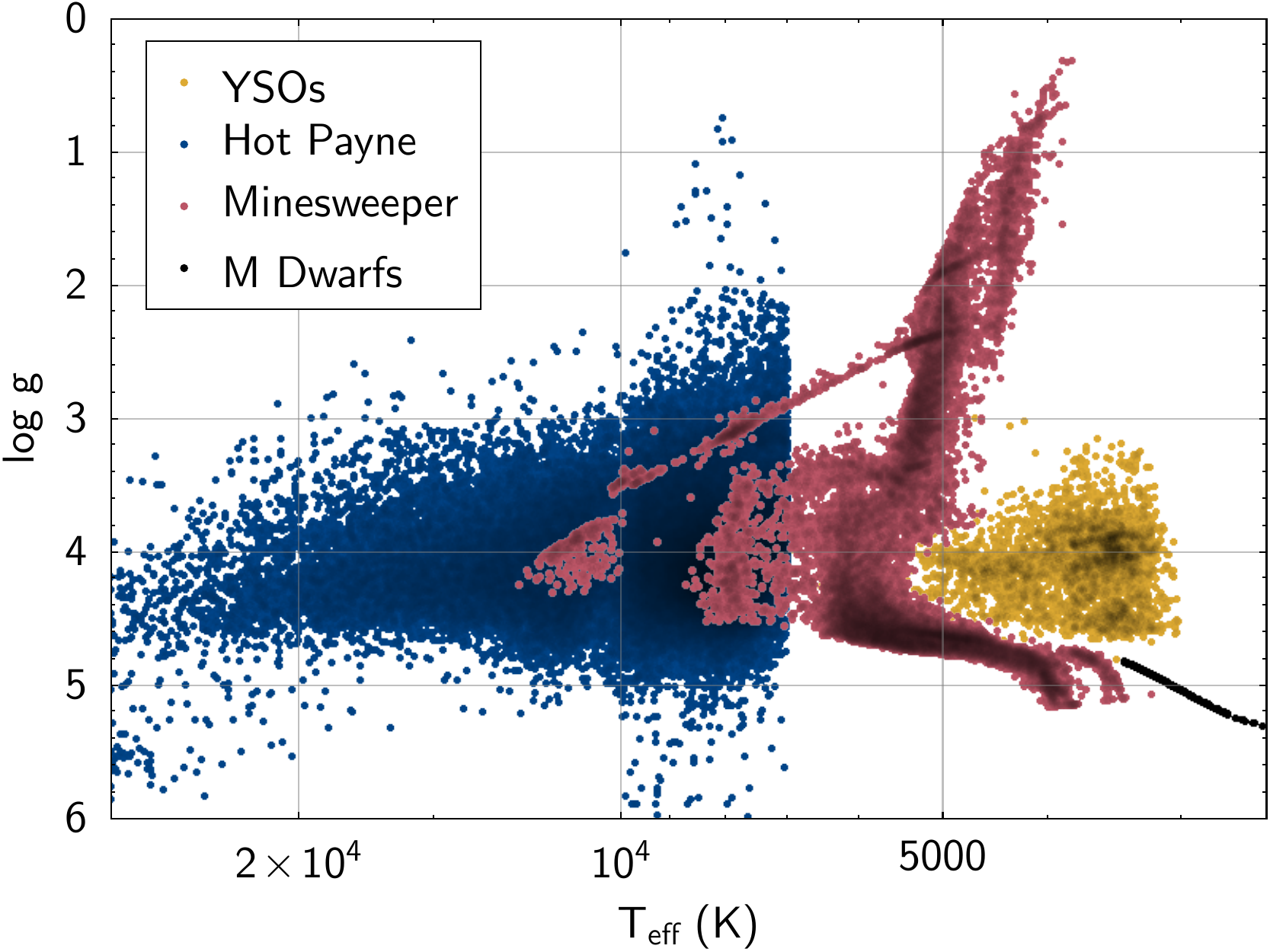}{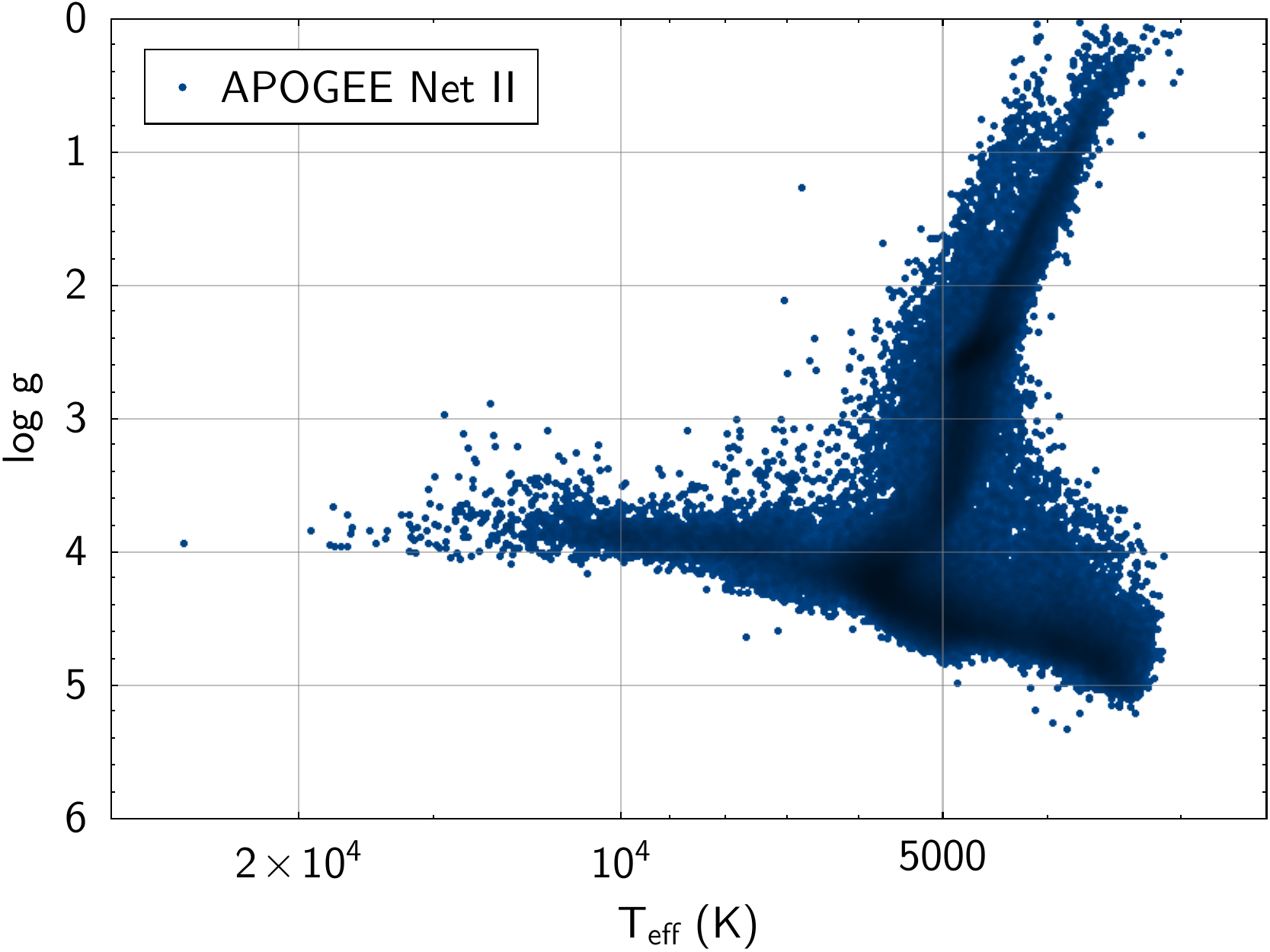}
\plottwo{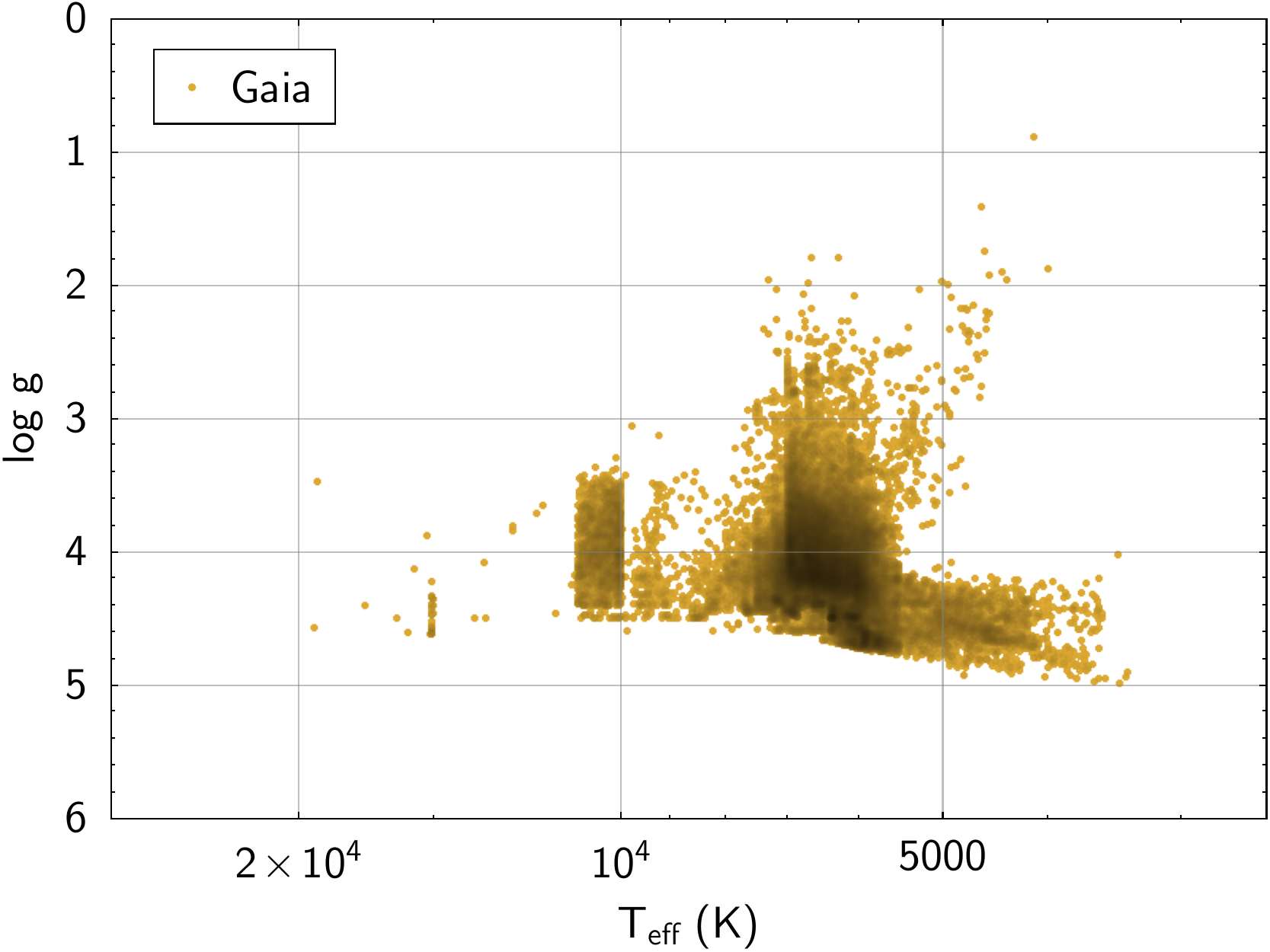}{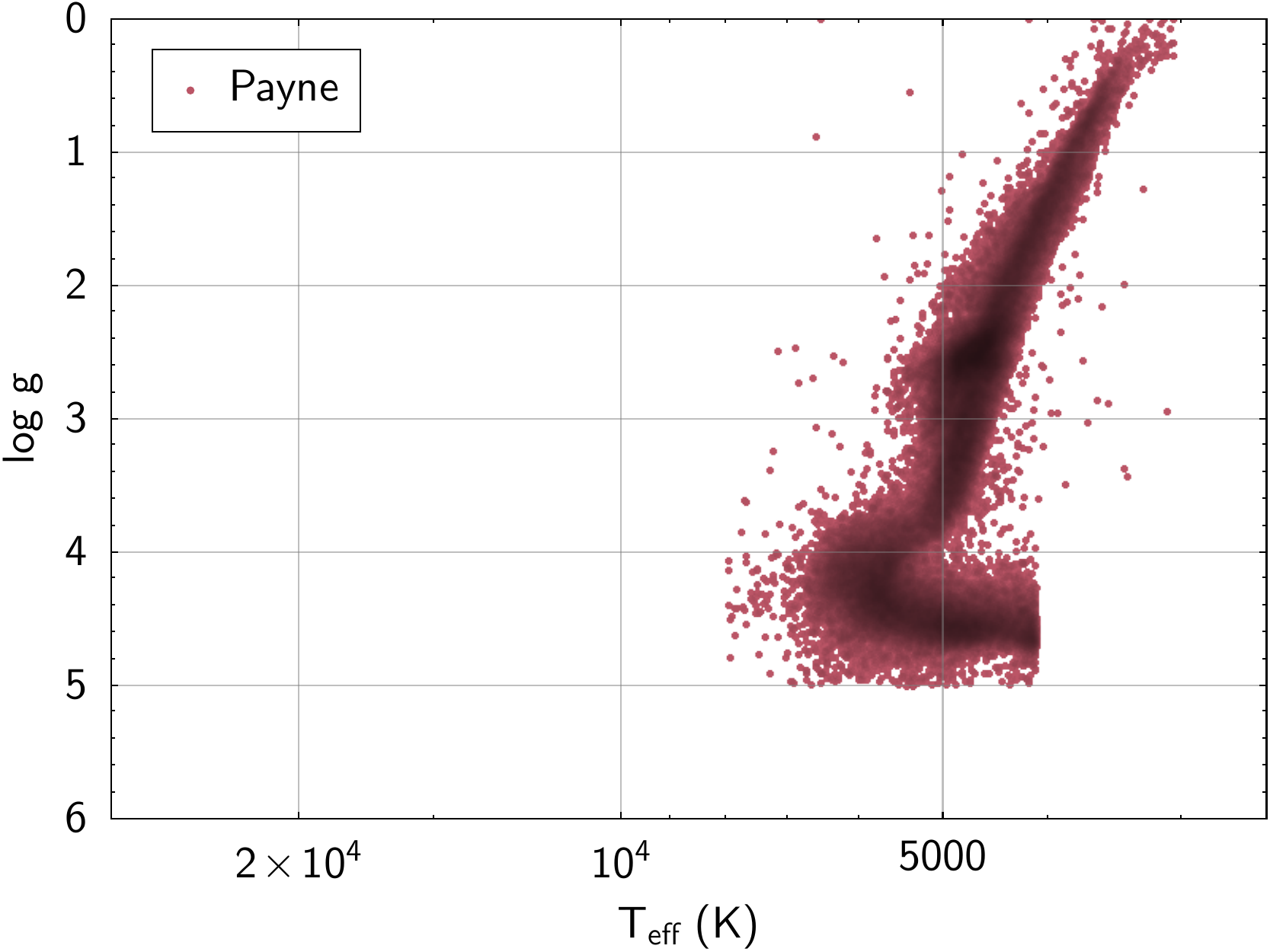}
\caption{Subsets of the different samples of the labelled data used in training.
\label{fig:trainingdetail}}
\end{figure*}

In Figure \ref{fig:trainingdetail} we show the location on the Kiel diagram of various subsets that were used to train BOSS Net. Most of them are located in well-isolated corners of the parameter space, which provides safeguards against the model predictions being systematically distorted in the overlap. However, among red giants and intermediate mass main sequence stars, there are 4 subsets, APOGEE Net, The Payne, Gaia, and MINESweper, which do have some overlap.

We thus compare labels produced by these pipelines in Table \ref{tab:comp}, to ensure they are sufficiently self-consistent. We note that in training each star is assigned only one label, but the comparison is done if independent measurements from both pipelines are available. APOGEE Net and The Payne have good agreement between each other; this is to be expected, since APOGEE Net initially was based on The Payne.

Stellar parameters produced by Gaia photometric pipeline have substantial systematic offset with respect to these two pipelines with respect to the red giants, but for sources with \logg$>$3.5 (since the vast majority of all sources for which labels have been adopted from Gaia are dwarfs), there is a much better agreement for both \teff\ and \logg, with only a slight scatter in [Fe/H].

MINESWeeper parameters are primarily computed for metal poor stars of the halo, which have not been observed with APOGEE, however, its catalog also includes a small subset of other sources that have been observed with both instruments. These sources are almost exclusively found on the red giant branch. Although there is a systematic offset in log \teff\ of 0.01 dex between the MINESWeeper model vs. ANet and The Payne which propagates to no nonlinear systematic differences in \logg, inflating the scatter, there is nontheless a good agreement between the two sets of labels.

Finally, we compare the outputs of the final model of BOSS Net to the labels in each of the subsets. In general, the performance across them is reasonably consistent, although some do have more scatter, for example, \logg\ measurements of YSOs, since the labels were derived photometrically, they were somewhat less precise than the spectroscopic labels in other subsets. Gaia [Fe/H] also appear to have more scatter relative to the predictions than other subsets as XP spectra from which they are derived has very low resolution, and may not have as much sensitivity towards very subtle changes. Finally, the parameters of subdwarfs and white dwarfs have sizeable scatter relative to the predictions because these spectra tend to have low SNR.

\begin{deluxetable*}{cccccccc}[!ht]
\tablecaption{Comparison of scatter between labels from different subsets.
\label{tab:comp}}
\tabletypesize{\scriptsize}
\tablewidth{\linewidth}
\tablehead{
 \colhead{Comparison} &
 \colhead{$\sigma$ \logg} &
 \colhead{$\sigma$ log \teff} &
 \colhead{$\sigma$ [Fe/H]} &
 \colhead{$\Delta$ \logg} &
 \colhead{$\Delta$ log \teff} &
 \colhead{$\Delta$ [Fe/H]} &
 \colhead{$N_{\rm star}$} 
 }
\startdata
ANet II---The Payne & 0.10 & 0.011 & 0.028 & -0.015 & -0.0002 & -0.0039 & 214725 \\
ANet II---Gaia (\logg$>3.5$) & 0.14 & 0.017 & 0.18 & -0.002 & -0.0027 & 0.069 & 206968 \\
The Payne---Gaia (\logg$>3.5$) & 0.14 & 0.015 & 0.18 & 0.022 & 0.0020 & 0.056 & 45442 \\
MINESweeper--- The Payne & 0.22 & 0.010 & 0.11 & -0.034 & 0.0075 & -0.075 & 2573 \\
MINESweeper--- Anet II & 0.27 & 0.012 & 0.09 & -0.007 & 0.007 & -0.073 & 4805 \\
MINESweeper--- Gaia (\logg$>3.5$) & 0.24 & 0.016, 0.36 & -0.10 & 0.0059 & 0.21 & 14253 \\
ANet II---Hot Payne & 0.35 & 0.043 & --- & -0.035 & 0.0049 & & 4836\\
\hline
\multicolumn{8}{c}{Scatter between the final model and the labels from different subsets\tablenotemark{$^a$}} \\
\hline
ANet II & 0.114 & 0.008 & 0.045 & 0.002 & -0.0004 & 0.0014 & 9356 \\
The Payne & 0.141 & 0.016 & 0.046 & 0.011 & -0.0005 & 0.0036 & 7501 \\
Gaia & 0.127 & 0.011 & 0.174 & 0.028 & 0.0012 & 0.0166 & 20412 \\
Hot Payne & 0.10 & 0.020 & 0.094 & -0.003 & 0.0007 & 0.0134 & 12514 \\
YSOs & 0.176 & 0.0134 & 0.028 & -0.037 & 0.0003 & 0.0012 & 317 \\
MINESweeper & 0.123 & 0.021 & 0.080 & 0.075 & 0.0013 & -0.0637 & 809 \\
Subdwarfs & 0.18 & 0.036 & --- & -0.022 & -0.0027 & --- & 91 \\
White dwarfs & 0.30 & 0.048 & --- & 0.050 & 0.0013 & --- & 428 \\
M dwarfs & 0.039 & 0.008 & --- & 0.017 & -0.0023 & --- & 773 \\
\hline
\multicolumn{8}{c}{Scatter between the BOSS \& LAMOST predictions for the same stars\tablenotemark{$^a$}} \\
\hline
BOSS---LAMOST & 0.105 & 0.011 & 0.104 & 0.026 & 0.0007 & -0.017 & 65013 \\
\enddata
\tablenotetext{a}{Reported only for the withheld test set.}
\end{deluxetable*}

\section{An update to the APOGEE Net}\label{sec:apogee}

\begin{deluxetable}{ccl}[!ht]
\tablecaption{Stellar properties from APOGEE DR18 spectra 
\label{tab:apogee}}
\tabletypesize{\scriptsize}
\tablewidth{\linewidth}
\tablehead{
 \colhead{Column} &
 \colhead{Unit} &
 \colhead{Description}
 }
\startdata
APOGEE\_ID & & APOGEE unique identifier \\
RA & deg & Right ascention in J2000 \\
Dec & deg & Declination in J2000 \\
log \teff & [K] & Effective temperature \\
$\sigma$ log \teff & [K] & uncertainty in log \teff \\
log g &  & Surface gravity \\
$\sigma$ log g & & Uncertainty in log g \\
$\left[\rm{Fe/H}\right]$&  & Metallicity \\
$\sigma$ [Fe/H] & & Uncertainty in [Fe/H] \\
\enddata
\end{deluxetable}

While APOGEE Net I and II \citep{olney2020,sprague2022} have performed well, they suffered from some limitations, such as: 
\begin{itemize}
    \item The original training set did not include brown dwarfs, as originally such sources were only rarely observed, but became increasingly more common in the trainig set. As their \teff\ is well outside the parameter range covered by ANet I \& II, this created erroneous for these type of sources without an ability to filter them out in any manner other than through photometeric cuts
    \item ANet II has expanded to include OB stars, but while it had produced reasonable \teff\, it struggled to provide a spread in \logg. Furthermore, the reliance of the model on photometry as metadata resulted in degraded performance in the areas of high extinction
    \item The distribution of \logg\ in coeval PMS stars did not match the expected slope based on the isochrones.
\end{itemize}
Taking advantage of improvement of labels that can be used for training offered through BOSS Net, we provide an updated APOGEE Net model.

\subsection{Labels}

We have reassembled the labels used to train APOGEE Net III.

\begin{itemize}
    \item We have used the parameters from APOGEE Net II for 101,152 stars, the remaining sources have relied on other catalogs.
    \item Although in most cases, their labels have not been substantively altered from APOGEE Net II, 196,080 stars were taken from BOSS Net labels derived either from BOSS or LAMOST spectra.
    \item 99,335 stars were taken from the original Payne catalog \citep{ting2019} to prevent degradation of labels.
    \item 455 stars were crossmatched from the zeta Panye catalog \citep{straumit2022}
    \item 6832 stars were cool dwarfs interpolated using relations described in Section \ref{sec:mdwarfs}
    \item 5638 stars were taken from from the Extended Stellar Parametrizer for Hot Stars catalog in Gaia DR3 \citep{fouesneau2022}. These sources appear to have reliable \teff, but there is a significant scatter in \logg. To improve these labels, we use a combination of these \teff s and photometrically derived radii that were reported by Gaia DR3, and interpolate them over the MIST isochrones to derive \logg. Additionally, younger massive stars that were targeted by the ABYSS program \citep{kounkel2023} were explicitly assumed to be main sequence in deriving their \logg.
    \item Pre-main sequence stars had their labels updated as described in Section \ref{sec:yso}. Further changes were required, however. Originally, ANet II has derived parameters for YSOs with \teff$>$3000 K. This limit was sufficient for both LAMOST and BOSS -- the faint limit of the targets, combined with the large distance to even the closest star forming regions, combined with extinction that is typical in young stars did not have significant census of cooler YSOs. On the other hand, APOGEE does appear to contain a number of young brown dwarfs, since it is a NIR specrograph, observing a number of YSOs during the era of the survey with longer exposure times. Without accounting for these sources in the labeled data, the model accurately identifies them to be cooler (albeit with a slight systematic offset of $\sim$100 K), but places them at wrong \logg\, forcing them onto the main sequence. 
    
    All likely YSOs were identified from a cross-match to various catalogs \citep[e.g.,][]{mcbride2021}, as well as from examining HR and Kiel diagrams. For the purposes of \logg\ determination, \teff was held to be at 3000 K, since MIST isochrones \citep{choi2016} do not extend to cooler stars. However, the \teff\ labels that were used for training the model were derived from SED fitting using BT-Settl synthetic templates \citep{allard2011}. In total, the training set consists of 8154 YSOs cooler than 7000 K, of which 525 have \teff$<$3000 K.
\end{itemize}

\subsection{Model}

The updated model was based on BOSS Net, keeping the same architecture and training method. Unlike APOGEE Net II, it no longer relied on photometry being passed as metadata to the model, since the labels for massive stars which previously needed it have been improved. The model predicts only 3 parameters, \teff, \logg, and [Fe/H], as RV determination is well-handled by the default APOGEE pipeline.

The model underwent fine-tuning in a manner consistent with the methods described earlier in the paper. Each stars loss is weighted according to the inverse of the probability density function of the parameter space, calculated via a KDE. Additionally, we apply further weight to the loss of stars in key astronomically significant regions of the parameter space, ensuring comprehensive modeling across the entire space.

\subsection{Results}
\begin{figure}
\epsscale{1.2}
\plotone{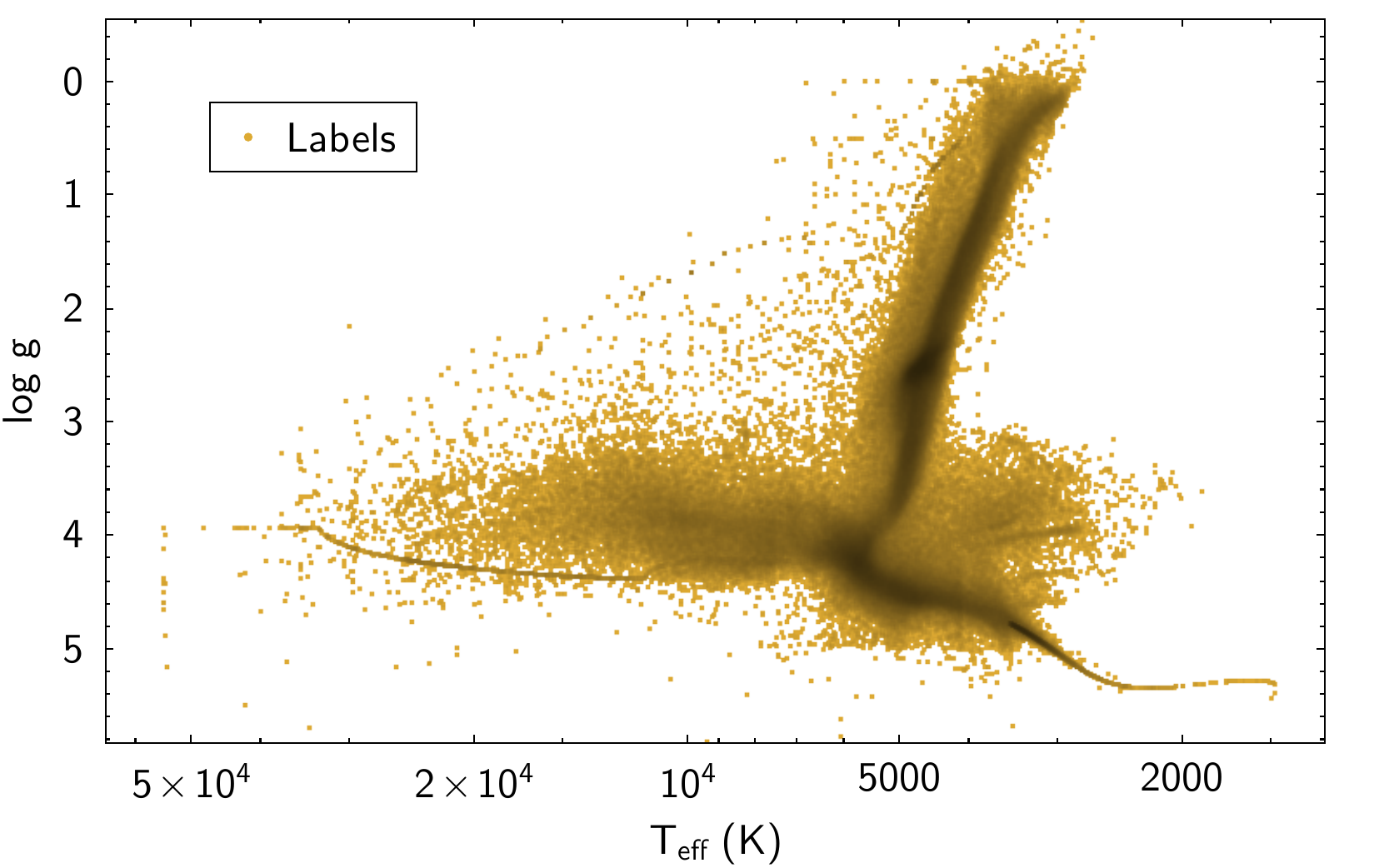}
\plotone{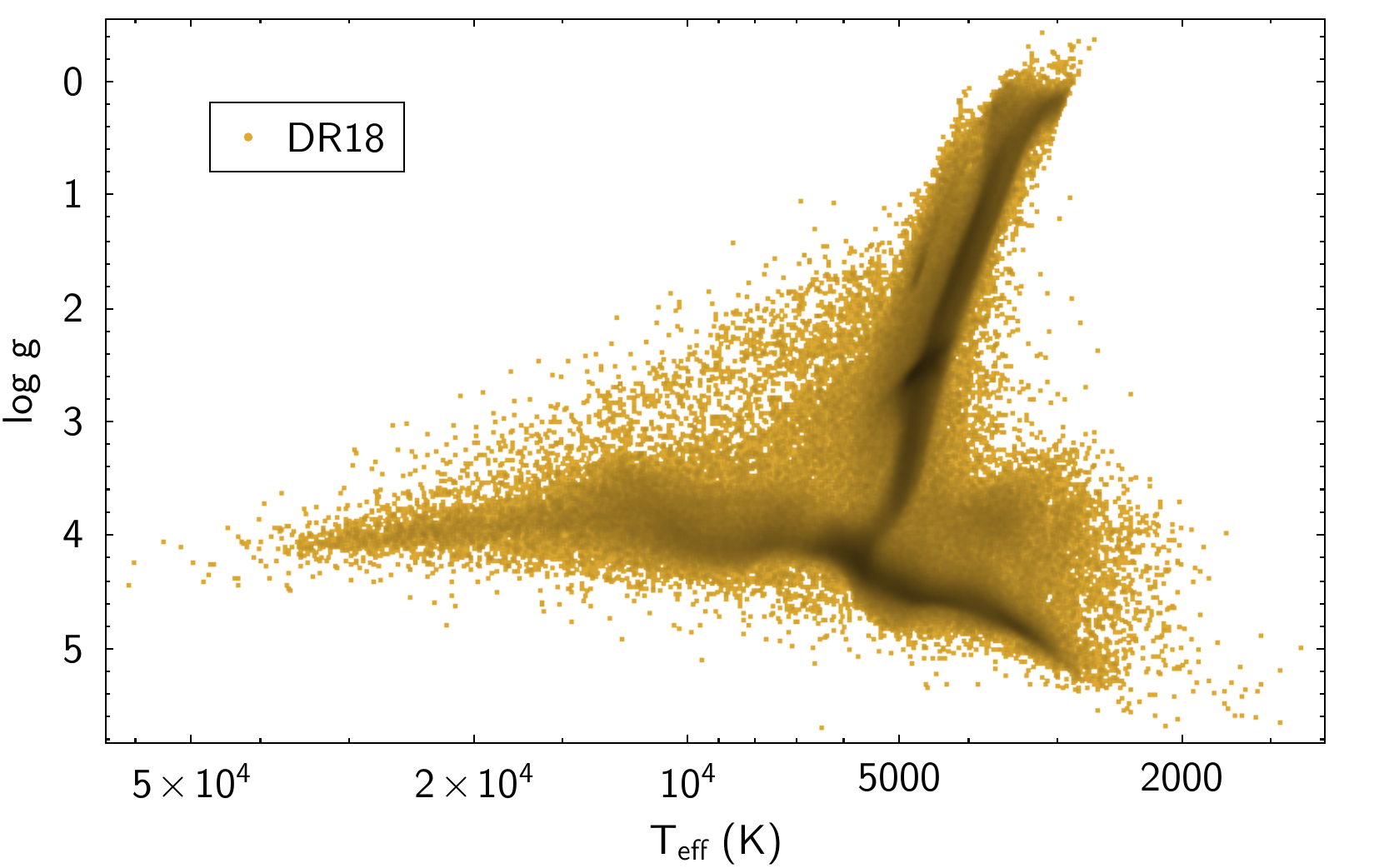}
\plotone{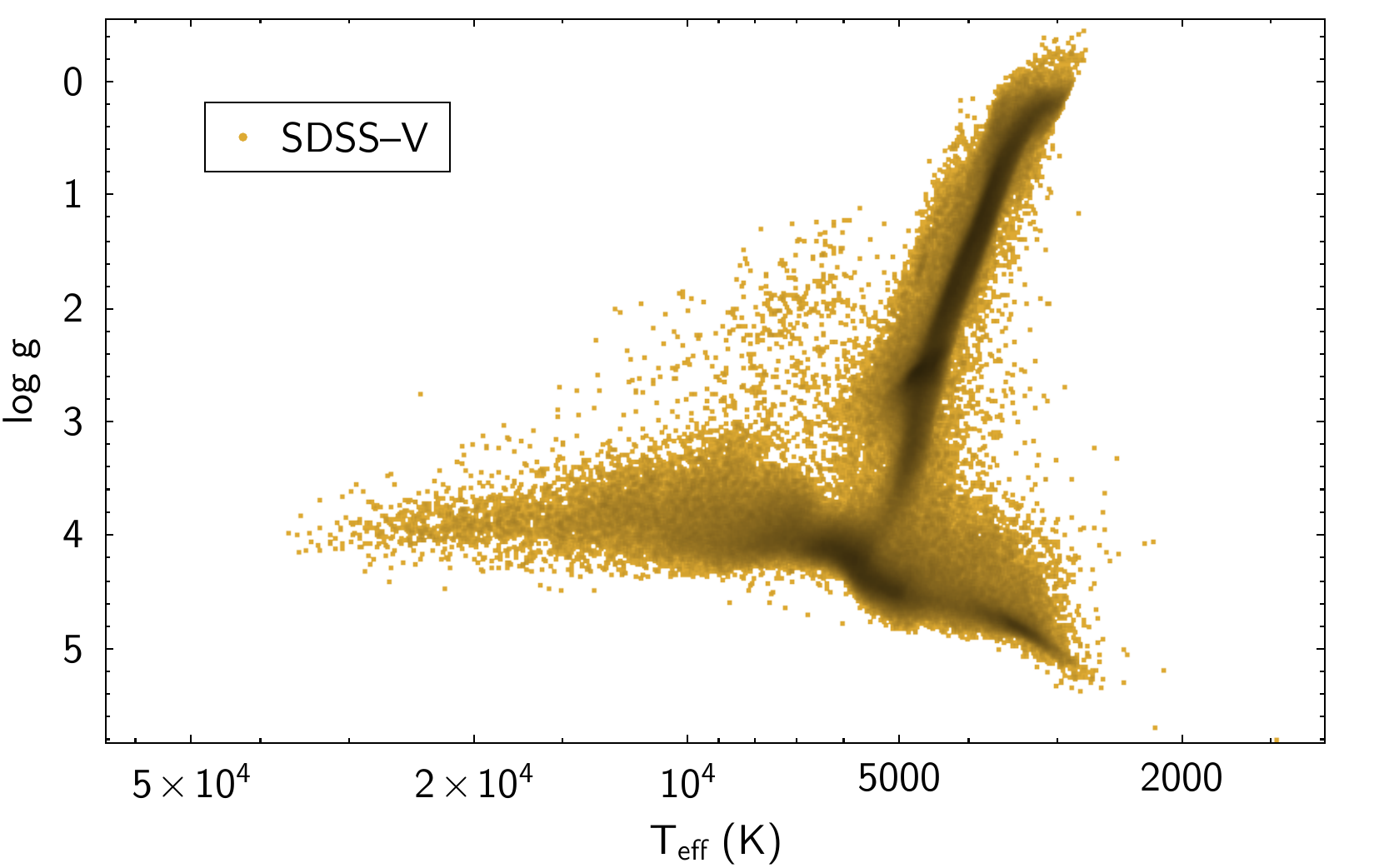}
\caption{Top: sample of labels used to train APOGEE Net. Bottom: predicted parameters for SDSS I-IV DR18 and for SDSS-V APOGEE data.
\label{fig:anet}}
\end{figure}

\begin{figure}
\epsscale{1.2}
\plotone{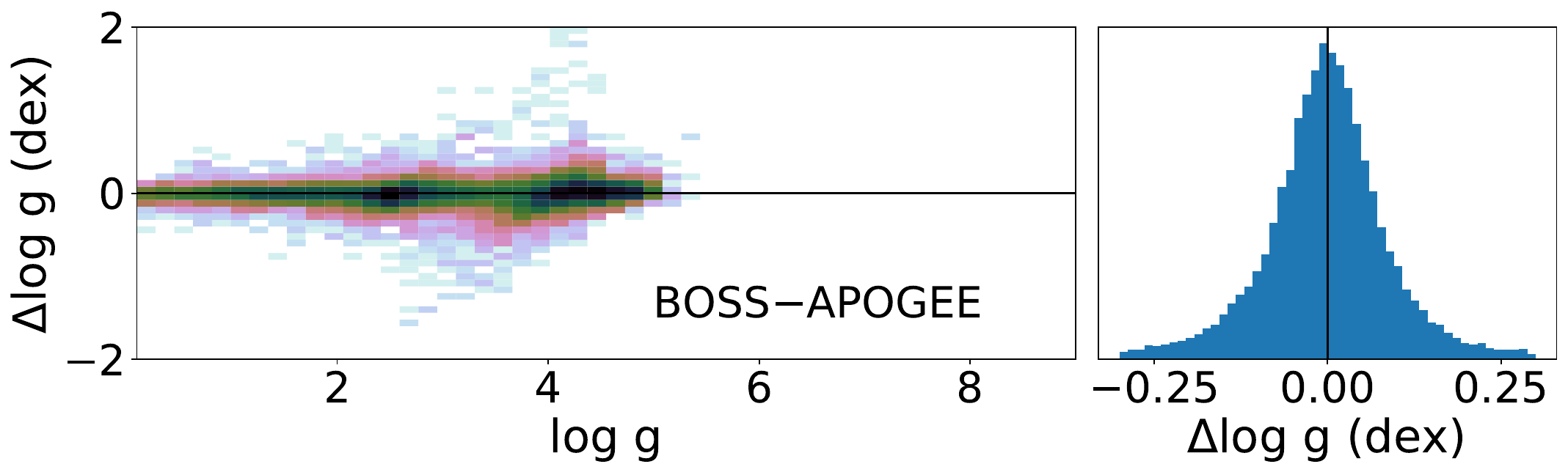}
\plotone{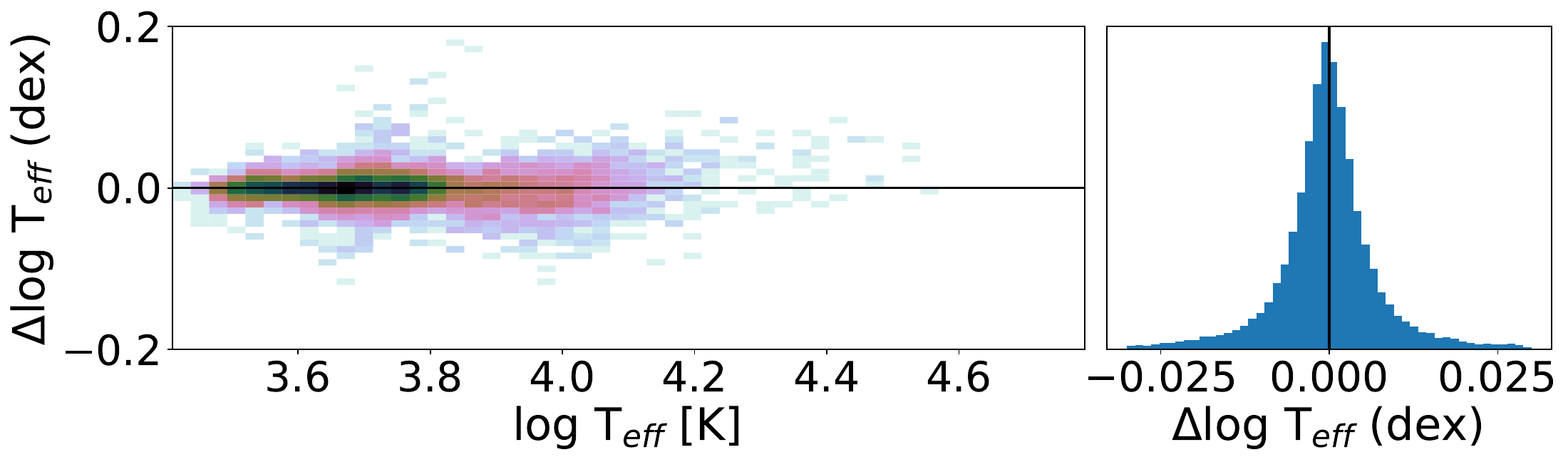}
\plotone{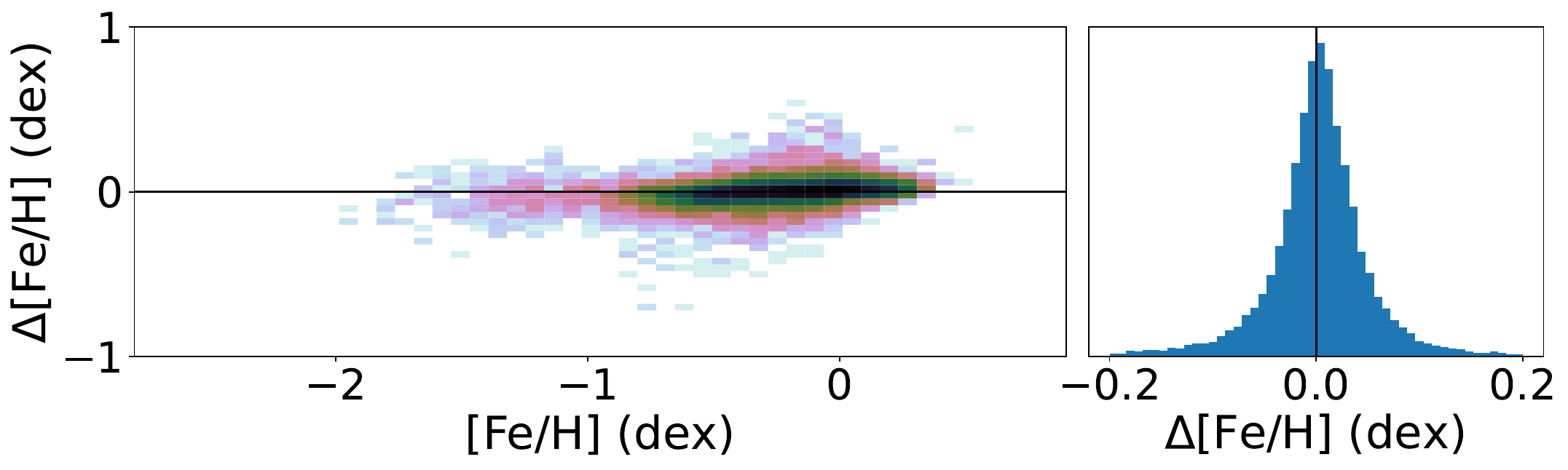}
\caption{Comparison between the derived parameters from APOGEE Net and BOSS Net (including both BOSS \& LAMOST spectra) of the sames sources, using the withheld test sample not used in training.
\label{fig:test3}}
\end{figure}

We evaluate the resulting model both for the legacy SDSS I-IV data (Table \ref{tab:apogee}), as well as SDSS-V data (which will be made available in subsequent data releases). The parameters are presented in Figure \ref{fig:anet}. Overall, the performance is consistent between APOGEE Net II and APOGEE Net III, with two notable improvements. The model is now capable of producing reliable parameters for stars with \teff$<$3000 K. Additionally, there is greater sensitivity to \logg\ at \teff$\sim$10,000 K than what has been achieved previously.

The performance between APOGEE Net and BOSS Net is similarly comparable: when both optical and IR spectra are available for the same sources, typical scatter between the measurements can be characterized by a Gaussian with the width of 1.1$\sigma$ for all of the parameters (Figure \ref{fig:test3}). At \teff$>$20,000 K, BOSS Net is producing somewhat more optimal \logg, for which there are multiple reasons. First, BOSS has observed significantly larger number of such sources for which labels are available. Additionally, optical spectra contain larger number of spectral features associated with OB stars than what is available in the H band.

Finally, we examine the sample differences between DR18 and SDSS-V sources that have been observed so far. The primary difference has been in pre-main sequence stars: SDSS-V lacks extremely young YSOs in the APOGEE sample, as due to the targeting strategy initial observations pointed specifically at several notable star forming regions have preferentially observed low mass YSOs with BOSS. Additionally, as previously mentioned, SDSS-V has shorter exposure times to enable the large volume of the survey, thus these YSOs do not reach as cool \teff\ as in the legacy era. There are also some differences in the targeted blue and yellow supergiants, although in both cases they are somewhat rare in comparison to BOSS data.

\end{appendix}

\software{TOPCAT \citep{topcat}, PyTorch \citep{pytorch}, BOSS Net \citep{bossnet}}

\begin{acknowledgements}

The authors thank Chao Liu, Jiadong Li, and Dan Qiu for their help in accessing LAMOST spectra. We also thank David Nidever and JJ Hermes for supplying some of the labels that have been helpful in improving the final model.

Funding for the Sloan Digital Sky Survey V has been provided by the Alfred P. Sloan Foundation, the Heising-Simons Foundation, the National Science Foundation, and the Participating Institutions. SDSS acknowledges support and resources from the Center for High-Performance Computing at the University of Utah. The SDSS web site is \url{www.sdss5.org}.

SDSS is managed by the Astrophysical Research Consortium for the Participating Institutions of the SDSS Collaboration, including the Carnegie Institution for Science, Chilean National Time Allocation Committee (CNTAC) ratified researchers, the Gotham Participation Group, Harvard University, Heidelberg University, The Johns Hopkins University, L'Ecole polytechnique f\'{e}d\'{e}rale de Lausanne (EPFL), Leibniz-Institut f\"{u}r Astrophysik Potsdam (AIP), Max-Planck-Institut f\"{u}r Astronomie (MPIA Heidelberg), Max-Planck-Institut f\"{u}r Extraterrestrische Physik (MPE), Nanjing University, National Astronomical Observatories of China (NAOC), New Mexico State University, The Ohio State University, Pennsylvania State University, Smithsonian Astrophysical Observatory, Space Telescope Science Institute (STScI), the Stellar Astrophysics Participation Group, Universidad Nacional Aut\'{o}noma de M\'{e}xico, University of Arizona, University of Colorado Boulder, University of Illinois at Urbana-Champaign, University of Toronto, University of Utah, University of Virginia, Yale University, and Yunnan University.

Guoshoujing Telescope (the Large Sky Area Multi-Object Fiber Spectroscopic Telescope LAMOST) is a National Major Scientific Project built by the Chinese Academy of Sciences. Funding for the project has been provided by the National Development and Reform Commission. LAMOST is operated and managed by the National Astronomical Observatories, Chinese Academy of Sciences.

This work has made use of data from the European Space Agency (ESA)
mission {\it Gaia} (\url{https://www.cosmos.esa.int/gaia}), processed by
the {\it Gaia} Data Processing and Analysis Consortium (DPAC,
\url{https://www.cosmos.esa.int/web/gaia/dpac/consortium}). Funding
for the DPAC has been provided by national institutions, in particular
the institutions participating in the {\it Gaia} Multilateral Agreement.

\end{acknowledgements}

\bibliographystyle{aasjournal.bst}
\bibliography{main.bbl}

\end{CJK*}
\end{document}